\documentclass[preprint]{aastex631}

\begin{document}

\title{Search for GeV gamma-ray emission from the possible TeV-bright red dwarfs with \textit{Fermi}-LAT}

\author[0000-0003-4916-4447]{CHEN HUANG}
\affiliation{School of Astronomy \& Space Science, Nanjing University, 163 Xianlin Avenue, Nanjing 210023, China}

\author[0000-0002-9392-547X]{XIAO ZHANG}
\affiliation{School of Astronomy \& Space Science, Nanjing University, 163 Xianlin Avenue, Nanjing 210023, China}

\affiliation{Key Laboratory of Modern Astronomy and Astrophysics, Nanjing University, Ministry of Education, Nanjing 210023, China}

\author[0000-0002-4753-2798]{YANG CHEN}
\affiliation{School of Astronomy \& Space Science, Nanjing University, 163 Xianlin Avenue, Nanjing 210023, China}
\affiliation{Key Laboratory of Modern Astronomy and Astrophysics, Nanjing University, Ministry of Education, Nanjing 210023, China}

\author[0000-0003-3717-2861]{WENJUAN ZHONG}
\affiliation{School of Astronomy \& Space Science, Nanjing University, 163 Xianlin Avenue, Nanjing 210023, China}

\correspondingauthor{Xiao Zhang \& Yang Chen}
\email{xiaozhang@nju.edu.cn; ygchen@nju.edu.cn}

\begin{abstract}

Red dwarfs have been suggested to be among the possible astrophysical species accelerating particles and emitting TeV $\gamma$-rays. As an effort to search for the GeV $\gamma$-ray counterparts of the suggested TeV emission from eight red dwarfs, we analyse the 0.2--500~GeV $\gamma$-ray emission of the regions covering them exploiting the $\sim$13.6 yr Pass 8 data of the \textit{Fermi} Large Area Telescope. A GeV $\gamma$-ray emission excess with significance of 3.8$\sigma$ is detected in the direction of the red dwarf V962\,Tau. This emission contains V962\,Tau in 1$\sigma$ error radius and is independent of the catalog source. However, the stellar flare scenario can hardly explain the total energy and lightcurve derived from the $\gamma$-ray emission in view of the spectral analysis. We also analyse the lightcurves in the positions of the eight red dwarfs and no time bin with significance $>$5$\sigma$ is found. Therefore, no significant emission from the red dwarfs could be concluded to be detected by \textit{Fermi}-LAT.

\end{abstract}


\section{Introduction} \label{sec:intro}

A large amount of studies in understanding the origin of cosmic rays (CRs) were made since their discovery \citep{2012ApJ...750....3A, 2022ApJ...931L..30B, 2023A&A...673A..75A}. Radio synchrotron radiation from supernova remnants (SNRs) proves that there are relativistic electrons accelerated in SNRs \citep{1954cpoc.conf..149G}, which means SNRs might be the potential source of Galactic cosmic rays. However, it is still ambiguous whether SNRs are the main accelerating place of Galactic cosmic rays, although pion-decay bumps, an important characteristic of proton-proton interaction shown in $\gamma$-ray spectra, have been detected in SNRs \citep{2013Sci...339..807A,2011ApJ...742L..30G}. If the Galactic supernovae solely sustain the observed cosmic-ray energy density, it needs an energy conversion efficiency, the energy ratio of cosmic rays to ejecta of supernova, as high as 10\% which is yet unproven \citep{2020pesr.book.....V, 2010ApJ...722L..58S, 2004MNRAS.353..550B}.  
Other sources than SNRs, such as stellar winds and flares, are believed to be also capable of accelerating particles to a high energy \citep{2010ARA&A..48..241B}. Some researches show that large flares in the Sun can produce $\gamma$-ray emission with energy up to tens of MeV by bremsstrahlung from energetic particles \citep{2019ApJ...877..145L} and $>100$~MeV by decay of neutral pions \citep{2020SoPh..295...18G, 2018ApJ...865L...7O, 2018PhRvD..98f3019T}.

It was recently reported that TeV emissions are detected in the direction of eight red dwarfs from 800~GeV to $\sim$20~TeV in the SHALON long-term observations and thus red dwarfs are considered as possible origin of CRs \citep{2019AdSpR..64.2585S}. Red dwarf is a kind of main sequence star, which is in the right bottom of H-R diagram, with a mass $\sim(0.075-0.5)M_\odot$ and a surface temperature 2500--5000K. Although red dwarfs have very low-masses compared to the Sun, the stellar activities of them, M dwarfs in particular, are more frequent and violent for the strong magnetic activity associated with their convective envelopes \citep{2017ApJ...834...92C}. It was shown that the occurrence frequency of flares from Sun-like stars or M dwarfs can be characterized by power-law $dN/dE\sim E^\alpha$ \citep{2011ASPC..448..197H,2021ApJ...910...41A,2019ApJS..241...29Y} and vary by orders of magnitude for different activity levels. The total output energy of one flare in red dwarf is $10^{32}$--$10^{35}$ erg, thousands of times higher than that of solar flare \citep{2017ApJ...849...36Y}. Hence it is plausible that red dwarfs which are more active than the Sun may be potential $\gamma$-ray sources. A powerful outburst from the flaring dwarf DG~CVn is detected by \textit{Swift} with associated optical emission \citep{Drake2014} and radio emissions \citep{2015MNRAS.446L..66F}. DG~CVn is observed in 0.1--100~GeV with \textit{Fermi}-LAT for the possible $\gamma$-ray emission \citep{2017MNRAS.467.4462L}. \cite{10.1093/mnras/stx2806} simulate the expected GeV $\gamma$-ray emission from DG CVn using the X-ray properties of the superflare. Recent observation of M dwarf TVLM 513-46546 shows a $\gamma$-ray pulse with a power-law index of $2.59\pm0.22$, and its $\gamma$-ray period is consistent with optical observations \citep{2020ApJ...900..185S}. It is known that red dwarfs are the most numerous stars and amount for more than 70\% of the Galactic stars. The total energy of stellar flares from all Galactic red dwarfs may be up to $\sim10^{51}$ erg \citep{2019AdSpR..64.2585S}, which is comparable to the estimation of the CR energy budget in the Galactic disk \citep{2011BRASP..75..323S}. Therefore, it is conjectured that red dwarfs maybe among the CR sources with energy up to $\sim10^{14}$ eV \citep{2021AN....342..342S}.

For exploring the ability of red dwarfs on accelerating particles, analysing the accompanied $\gamma$-ray emission is necessary.
In this study, we perform a detailed GeV $\gamma$-ray analysis of the eight red dwarfs which are reported as the TeV sources \citep{2021AN....342..342S} by using $\sim$13.6 yr \textit{Fermi}-LAT data. The sample of red dwarfs with the observational physical parameters are listed in Table \ref{table1}. We find that only V962\,Tau and GJ\,1078 may have the GeV emission. This paper is organized as follows. In Section \ref{sec:2}, we present the data analysis and the results. In Section \ref{sec:3} we discuss the possible association between the GeV source and V962\,Tau. Finally, a summary is provided in Section \ref{sec:4}.

\begin{table*}
 \caption{Parameters of the red dwarfs.}
  \centering
  \begin{tabular}{lccccccc}
    \hline
    Name & R.A. (J2000) & Dec (J2000) & $l$ & $b$ & Distance & Spectral type & Reference  \\
         & (hh mm ss) & (dd mm ss) & (degree) & (degree) & (pc) & & \\
    \hline
    V388 Cas  & 01 03 19.8350 & +62 21 55.8372 & 124.3105 & $-$00.4744 &   9.9 & M5V & 1, 2 \\
    V547 Cas  & 00 32 29.5078 & +67 14 09.1390 & 121.0955 & $+$04.4317 &   9.9 & M2V & 1, 2 \\
    V780 Tau  & 05 40 25.7311 & +24 48 07.8631 & 182.9119 & $-$03.1557 &  10.2 & M7V & 3, 4, 5 \\
    V962\,Tau  & 05 45 51.9441 & +22 52 47.3984 & 185.2040 & $-$03.1029 & 710.3 & ——  & 1 \\
    V1589 Cyg & 20 42 49.1546 & +41 22 59.9732 & 081.3674 & $-$00.5990 &  —— & M3.87 & 5, 6 \\
    GJ\,1078   & 05 23 49.0464 & +22 32 38.7532 & 182.7519 & $-$07.5756 &  24.9 & M4.5V & 3, 5, 7\\
    GJ 3684   & 11 47 05.1623 & +70 01 58.6481 & 130.7759 & $+$46.1028 &  25.9 & M4V & 5, 8 \\
    GL 851.1  & 22 12 06.4164 & +31 33 41.0764 & 087.3969 & $-$20.1010 &  41.1 & K5V & 1, 9 \\
    \hline
  \end{tabular}
  \label{table1}
  \begin{flushleft}
  \textbf{Reference.} (1) \cite{2020yCat.1350....0G}, (2) \cite{1991ApJS...77..417K}, (3) \cite{2014ApJ...784..156D}, (4) \cite{2014AJ....147...20N}, (5) \cite{2012yCat.1322....0Z}, (6) \cite{2015ApJS..220...16T}, (7) \cite{1995AJ....110.1838R}, (8) \cite{2015ApJ...812....3W}, (9) \cite{1986AJ.....92..139S}.
  \end{flushleft}
\end{table*}

\section{Data Analysis}
\label{sec:2}
\subsection{Analysis setup}
We analyse more than 13.6 years (from 2008-08-04 15:43:36 (UTC) to 2022-03-15 05:41:24 (UTC)) of Fermi-LAT Pass 8 SOURCE class (evclass=128, evtype=3) data with the software Fermitools 2.0.8\footnote{\url{https://fermi.gsfc.nasa.gov/ssc/data/analysis/software/}}. The regions of interest (ROIs) in our study are $15^{\circ}\times15^{\circ}$ in size, centered at each of the eight red dwarfs.

Firstly, the data selection is made with command \textit{gtselect} with the maximum zenith angle of 90$^{\circ}$. We apply command \textit{gtmktime} to the data with recommended filter string “(DATA\_QUAL \textgreater 0)\&\&(LAT\_CONFIG == 1)” for choosing good time intervals. The entire energy range from 0.2~GeV to 500~GeV is divided into 10 logarithmic bins per decade for counts cube and exposure cube. The appropriate Instrument Response Functions is “P8R3\_SOURCE\_V3” used for these data set, and the Galactic interstellar diffuse background emission model and isotropic background spectral template ``\textit{gll\_iem\_v07}” and ``\textit{iso\_P8R3\_SOURCE\_V3\_v1}” are applied, respectively. 
Then, we use the 4FGL-DR3 catalog, the \textit{Fermi}-LAT 12-year source catalog, and the two background models to define our source model by the user-contributed tool \textit{make4FGLxml.py}\footnote{\url{https://fermi.gsfc.nasa.gov/ssc/data/analysis/user/make4FGLxml.py}}, and the corresponding list of 4FGL sources within a radius of 25° centered at target source is obtained. 
After that we use the python module \textit{pyLikelihood} with the NEWMINUIT optimiser to perform the binned likelihood analysis and get the best-fit results. In this step, we just free the spectral parameters of the catalog sources within 5$^{\circ}$ from the ROI centers and the normalization of the two diffuse background components.
In addition, a python package FERMIPY \citep{2017ICRC...35..824W} is employed
\footnote{\url{https://fermipy.readthedocs.io/en/latest/}} (version 1.2) in the position fitting process and the lightcurve (LC) analysis.

\subsection{Results}
\label{sec:2.2}

For searching the GeV counterparts from the suggested TeV-bright red dwarfs listed in Table \ref{table1}, we generate their residual test-statistic (TS) maps in the energy range from 0.2~GeV to 500~GeV. The TS value for each pixel is evaluated by ${\rm TS}=2{\rm \ln}(L_1/L_0)$, where $L_0$ is the maximum likelihood of the null hypothesis and $L_1$ the maximum likelihood of the test model that a putative point source is located in this pixel.

\begin{figure*}
  \centering
	\includegraphics[width=8cm]{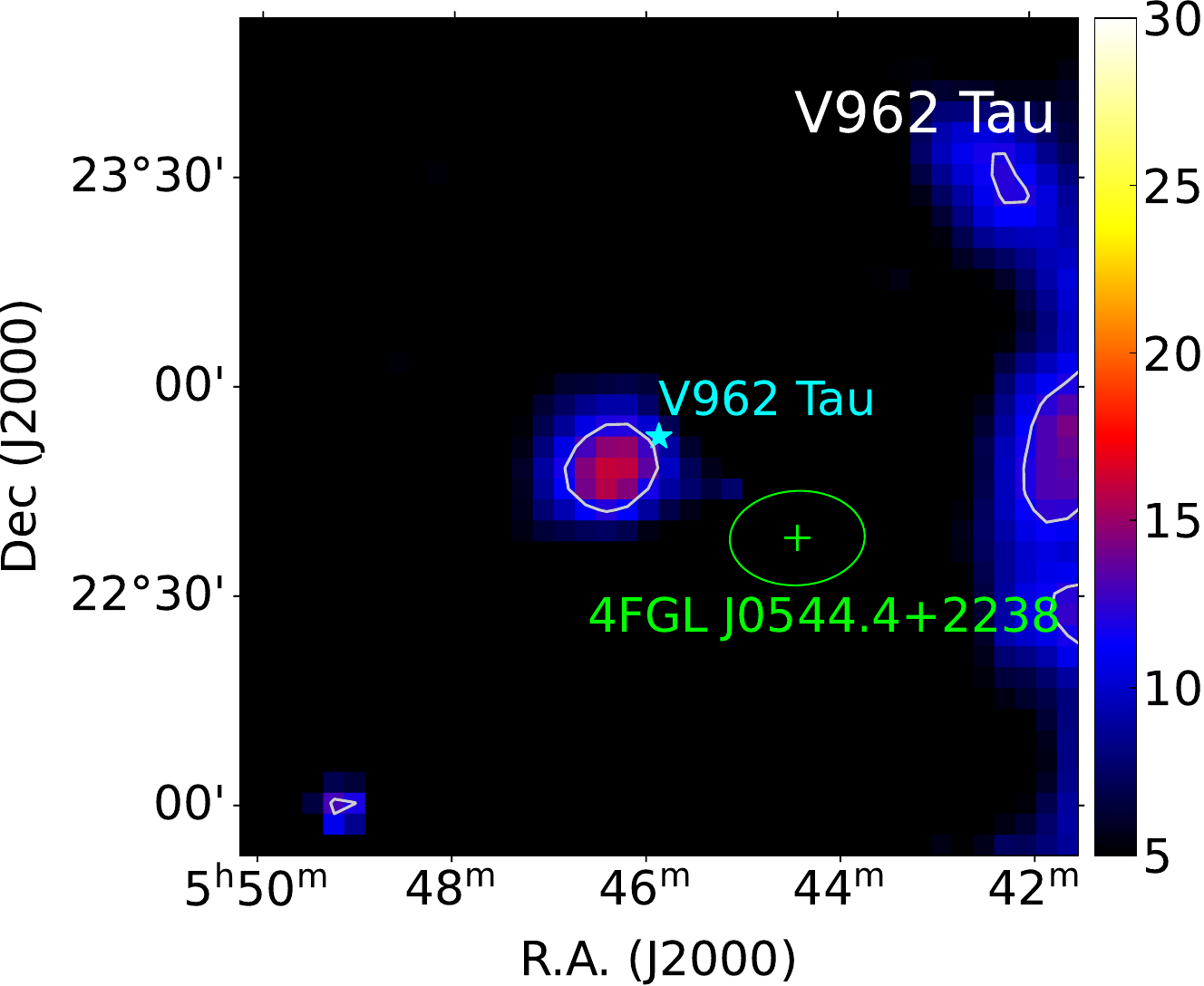}
	\includegraphics[width=8cm]{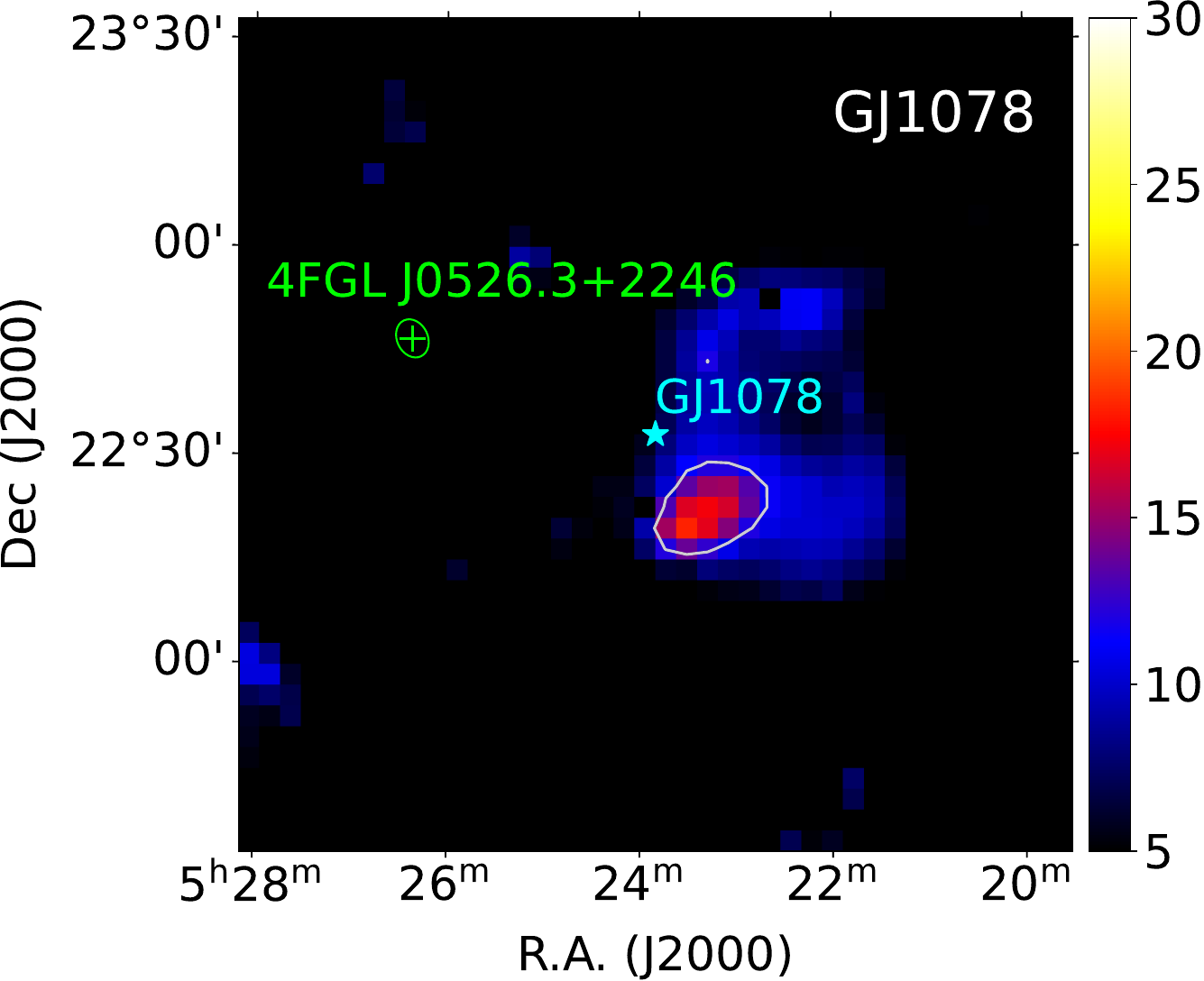}
  \caption{TS maps of 2$^{\circ}$ × 2$^{\circ}$ regions centered at V962~Tau~(\textit{Left}) and GJ~1078~(\textit{Right}) in the energy range of 1--500~GeV. The green pluses mark the positions of 4FGL-DR3 catalog sources with 95\% error ellipses shown in green ellipses. The cyan stars mark the positions of red dwarfs. The white contours show the emission excess with 3~$\sigma$ significance which is corresponding to ${\rm TS}=11.83$ with 2 degrees of freedom.}
  \label{fig:tsmap_2src}
\end{figure*}

Besides some nearby excesses, no $>3\sigma$ GeV emission is found coincident with any of the eight red dwarfs in our spatial analysis and all of them are outside 95\% error ellipses of nearby 4FGL catalog sources.
In order to investigate the possible association between the targets and the nearby excesses, we use the 1--500 GeV data with better angular resolution to generate the $2^{\circ}\times2^{\circ}$ TS maps where only the catalog sources and two diffuse backgrounds are modelled.
We found that there are $>3\sigma$ excess near V962~Tau and GJ~1078 with an angular separation of a few $0.1^{\circ}$, which are shown in Figure~\ref{fig:tsmap_2src}.
Thus the detailed analysis for V962~Tau and GJ~1078 is presented in the following.
For the remaining sources, the TS maps can be found in Appendix~\ref{appendix}.

\subsubsection{V962 Tau}
\label{sec:2.3}

\begin{figure*}
  \centering
	\includegraphics[width=8cm]{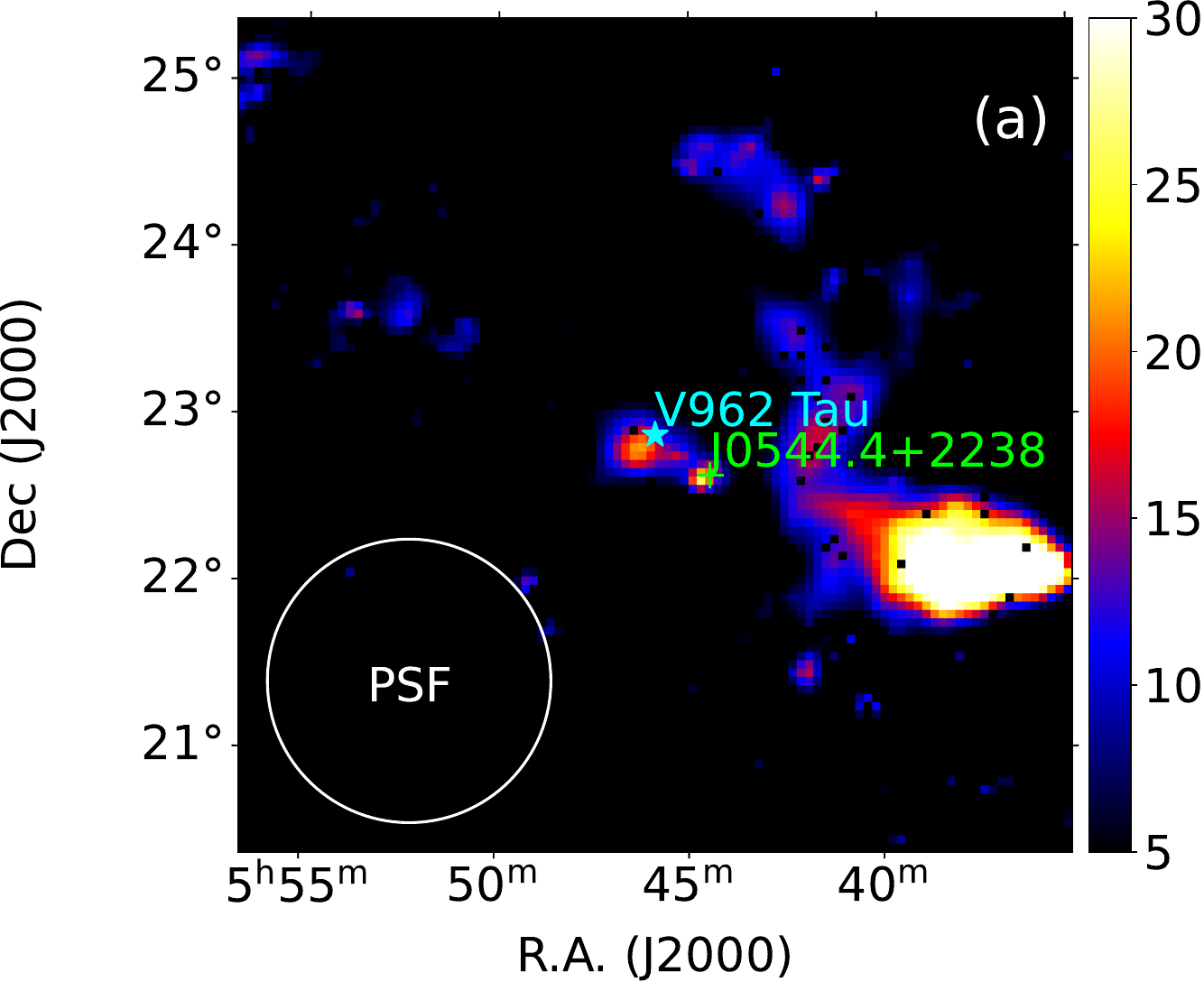}
	\includegraphics[width=8cm]{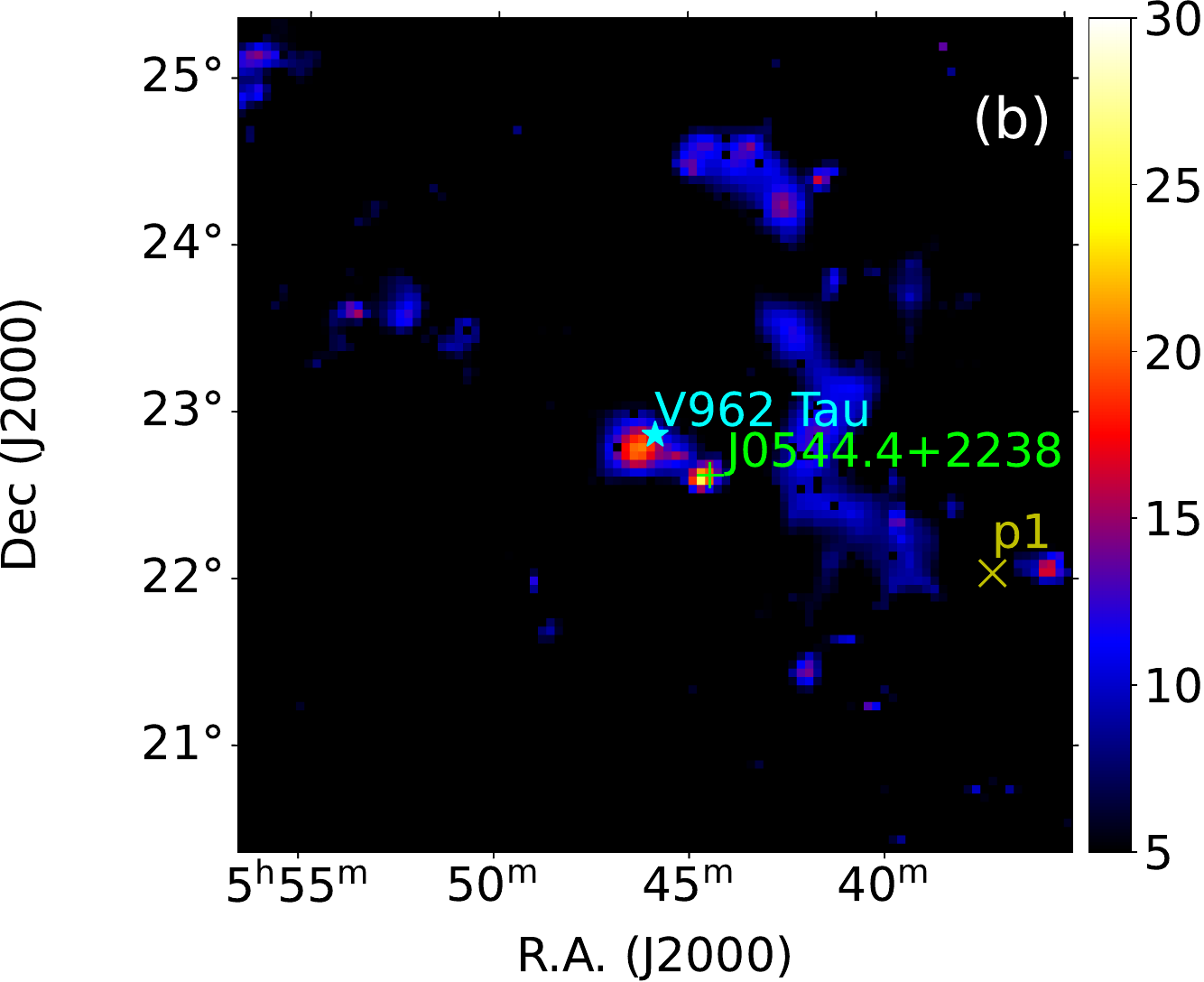}
	\includegraphics[width=8cm]{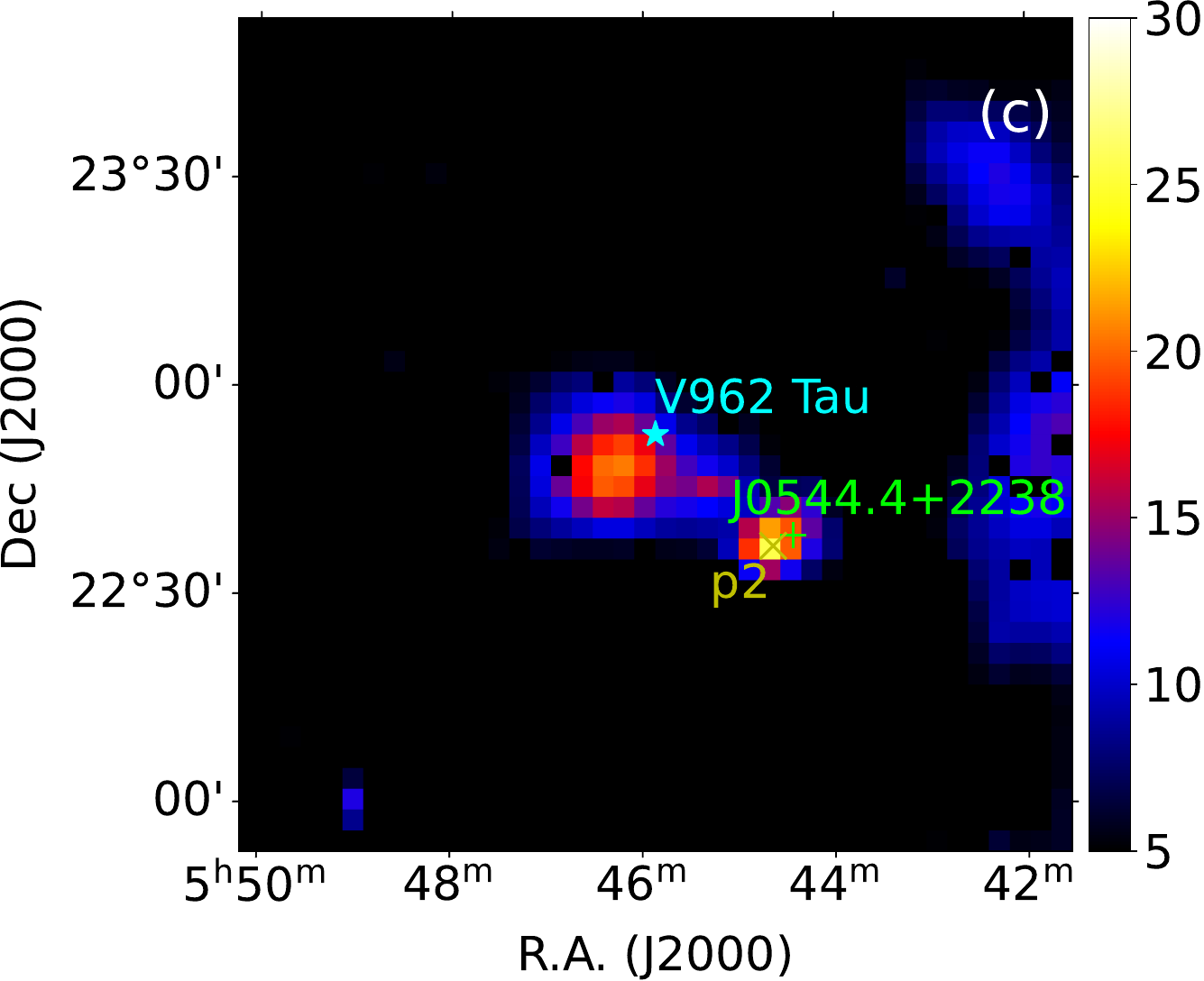}
	\includegraphics[width=8cm]{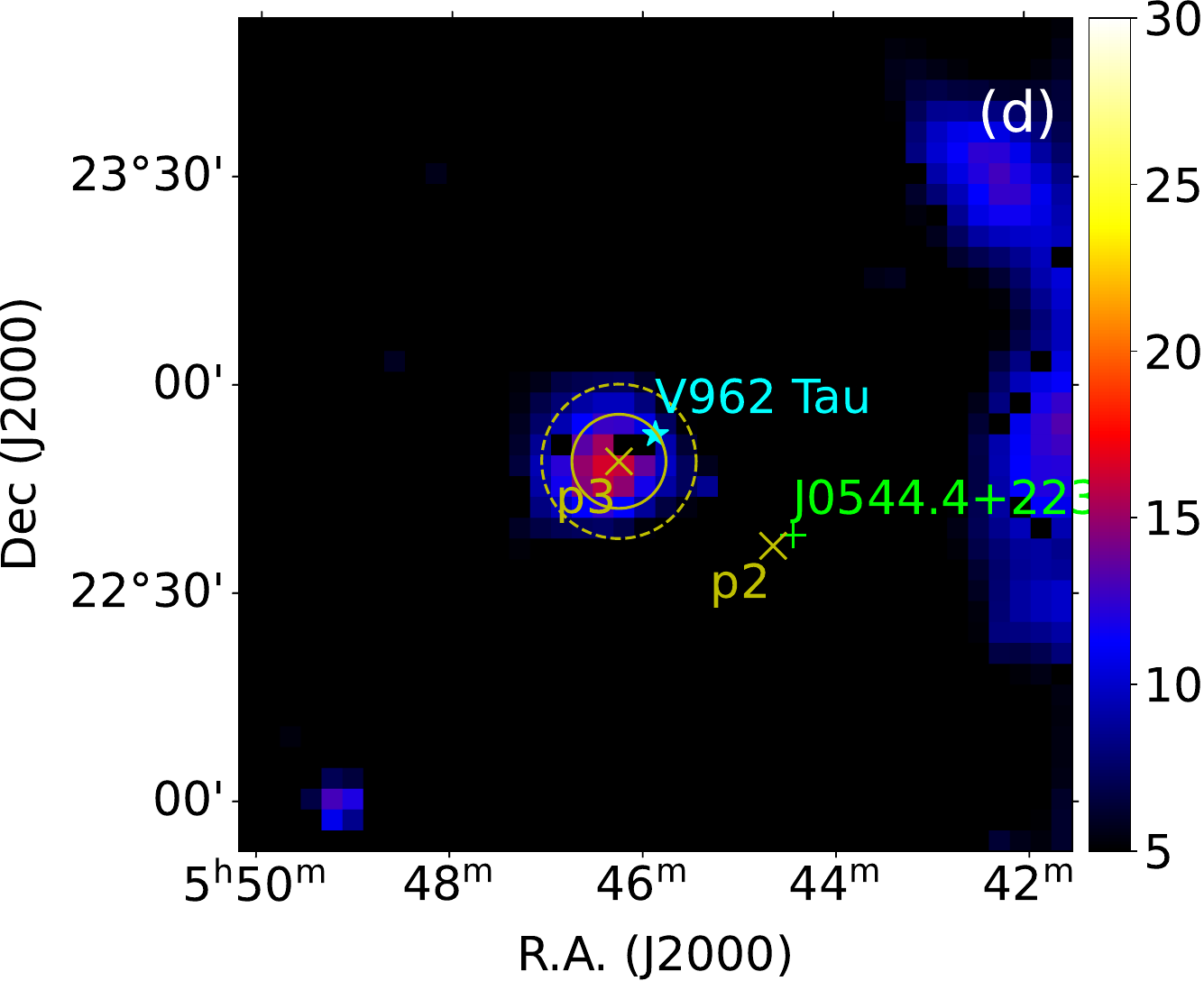}
  \caption{TS maps centered at V962\,Tau in the energy range of 1–-500~GeV. {\bf (a)} The $5^{\circ}\times5^{\circ}$ region by excluding 4FGL J0544.4+2238 from the source model. The white circle in the lower left shows the 68\% containment range of point spread function (PSF) of \textit{Fermi}-LAT at 1~GeV. {\bf (b)} Same as (a) but a point source p1 is included in the source model. {\bf (c)} Same as (b) but for a $2^{\circ}\times2^{\circ}$ region. {\bf (d)} Same as (c) but p2 is added to the source model. The yellow solid and dashed circles represent the $1\sigma$ and $2\sigma$ error radii of source p3, respectively.}
  \label{fig:tsmap_v962}
\end{figure*}

For V962\,Tau, there are some excess in the east of the field and a nearby catalog source 4FGL J0544.4+2238 in the southwest (see the left panel of Figure~\ref{fig:tsmap_2src}), which may affect the result. 
Thus, we firstly generated a $5^{\circ}\times5^{\circ}$ TS map in the energy range of 1--500~GeV by excluding 4FGL J0544.4+2238 from the source model.
As shown in Figure~\ref{fig:tsmap_v962}a, a significant excess is present in the nearby southwestern region of V962\,Tau.
To eliminate the influence from this source, we add a point source p1 with a simple power-law spectrum at the peak pixel (${\rm R.A._{J2000}}=84.285^{\circ}$, ${\rm Dec_{J2000}}=22.040^{\circ}$) to the source model and reproduce the TS map as shown in Figure~\ref{fig:tsmap_v962}b.
There are some residual emission with two peaks near V962\,Tau as shown in Figure~\ref{fig:tsmap_v962}c.
Due to the small separation between the two peaks, we use another point source p2 to model the residual emission and use the \textit{localize} method in FERMIPY to fit its position.
The best-fit position of p2 is ${\rm R.A._{J2000}}=86.161^{\circ}$, ${\rm Dec_{J2000}}=22.617^{\circ}$ with a $1\sigma$ error radius of $0.033^{\circ}$, very close to that of 4FGL J0544.4+2238, which means that p2 is precisely 4FGL J0544.4+2238.
After subtracting the contribution of p1 and p2, the residual TS map is displayed in Figure~\ref{fig:tsmap_v962}d, which still shows an excess to the southeast of V962~Tau with a peak TS value of $\sim$17.
Therefore, we add the third point source p3 and fit its position, resulting in ${\rm R.A._{J2000}}=86.562^{\circ}$, ${\rm Dec_{J2000}}=22.820^{\circ}$ with a $1\sigma$ error radius of $0.113^{\circ}$.
The corresponding TS value of p1, p2, and p3 are ${\rm TS_{p1}}=72.1$, ${\rm TS_{p2}}=22.3$, and ${\rm TS_{p3}}=18.9$, respectively.
To further check the significance of p3, we calculate the likelihood ratio of the 3-point-source model (p1+p2+p3, $L_{\rm 3ps}$) to the 2-point-source model (p1+p2, $L_{\rm 2ps}$) via ${\rm TS_{model}}=2{\rm \ln}(L_{\rm 3ps}/L_{\rm 2ps})$, obtaining ${\rm TS_{model}}=17.5$ that corresponds to a significance of $3.8\sigma$ above 1 GeV.
Although the significance of p3 is not too high, it may be the possible signal as the marginal detection.
Thus, we will adopt the 3-point-source model in the following analysis for V962\,Tau.

In the 3-point-source model, both p2 and p3 are in the vicinity of V962\,Tau with an angular distance of $0.389^{\circ}$ and $0.109^{\circ}$, respectively.
Due to the small $1\sigma$ statistical error radius ($\sim$0.03$^{\circ}$), however, V962\,Tau is obviously outside the $3\sigma$ positional uncertainty radius\footnote{The $2\sigma$ and $3\sigma$ error radius are calculated as 1.6407$r_{1\sigma err}$ and 2.2699$r_{1\sigma err}$ with 2 degrees of freedom \citep{1996ApJ...461..396M}.} of p2 even if the systematic error is taken into account\footnote{The total uncertainty radius is calculated as $r_{\rm tot}^2 = (1.06r_{\rm stat})^2+r_{\rm abs}^2$, where $r_{\rm abs}=0.0068^{\circ}$ \citep{4FGL-DR1}.}.
While V962\,Tau is located within the $1\sigma$ error radius of source p3 after considering the systematic error.
Compared with p2, therefore, source p3 is more likely related to V962\,Tau from the spatial distribution.

To further explore the property of source p3 and its relation to V962\,Tau, we perform a spectral analysis in 0.2--500~GeV. 
It is noted that the distance between p3 and Crab is about 2.8$^{\circ}$ while the 68\% containment range of PSF at energy 0.2~GeV is about 3$^{\circ}$, which may affect the flux below 1~GeV.
To reduce the influence from Crab' pulsar, we only use the off-pulse (0.45--0.85) interval of Crab in phase-space based on the ephemeris from \cite{2010ApJ...708.1254A} to calculate the energy flux between 0.2--500~GeV.
With a power-law spectrum, the photon index $\Gamma=3.08\pm0.08$ and the energy flux of $(7.59\pm 1.15) \times 10^{-12}\ {\rm erg\ cm^{-2}\ s^{-1}}$ are obtained in the global fitting.
Assuming a distance of 710~pc \citep{2020yCat.1350....0G}, the luminosity of p3 is $4.58\times10^{32}\ {\rm erg\ s^{-1}}$.
The spectral energy distributions (SEDs) of p3 are produced by the \textit{sed} method in FERMIPY in five logarithmically spaced energy bins based on the maximum likelihood analysis. 
During the fitting process, the free parameters only include the normalization parameters of the sources with the significance $\geq5\sigma$ within 5$^{\circ}$ from the ROI centers as well as the Galactic and isotropic diffuse background components, while all other parameters are fixed to their best-fit values in the global fitting. For these energy bins with TS~$\le4$, the 95\%-confidence-level upper limits are calculated.

\begin{figure}
    \centering
    \includegraphics[width=8.4cm]{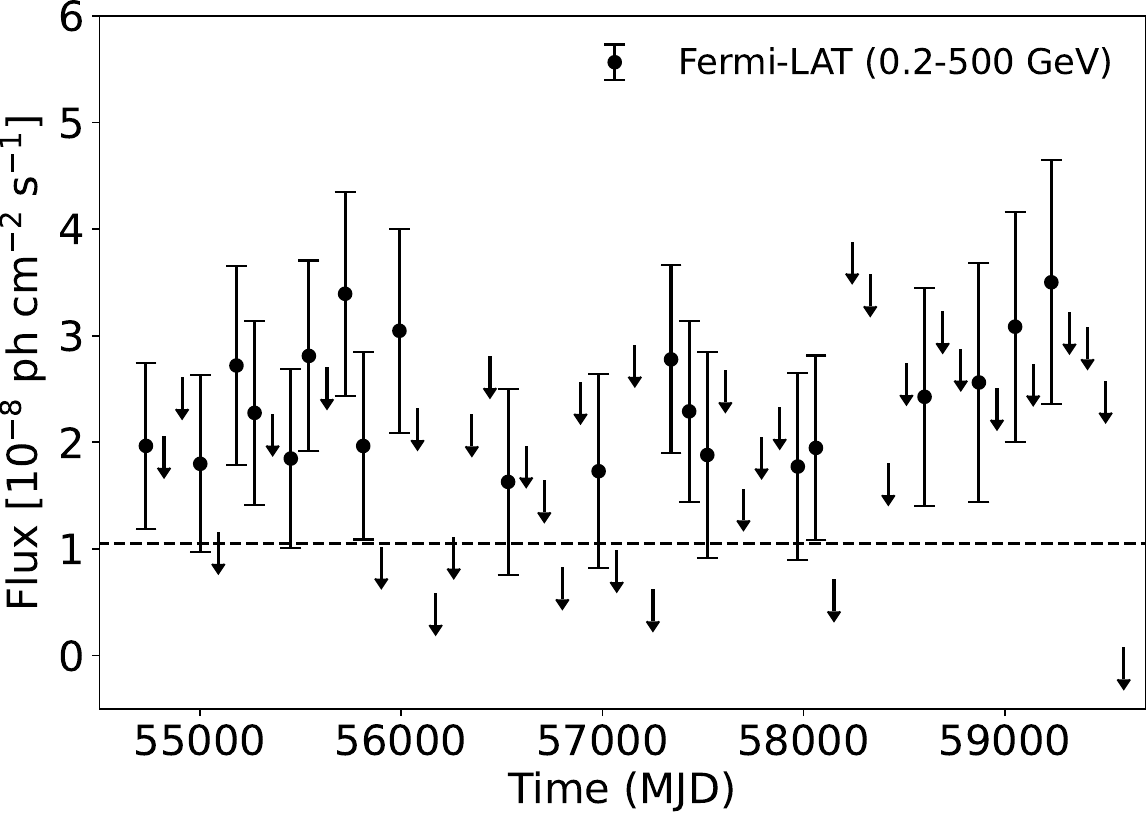}
  \caption{Three-month binned $\gamma$-ray lightcurve of p3 near V962\,Tau. The horizontal dashed line represents the constant flux. For the bins with TS~$\leq4$, the 95\% upper limits are presented.}
  \label{fig:lc_p3_v962}
\end{figure}

In order to examine the long-term variability of source p3, three-month binned lightcurve is constructed over the whole time and energy interval by using the LIGHTCURVE method in FERMIPY. 
The LC is shown in Figure~\ref{fig:lc_p3_v962}, in which for the bins with TS~$\leq4$, the 95\%-confidence upper limits are displayed.
According to the criterion in \citet{2FGL}, variable source can be identified at a 99\% confidence level if the variability index ($\rm TS_{var}$) is greater than 88.6 for the 55 time bins.
In view of ${\rm TS_{var}=59.5}$ obtained for p3 in this study, no significant $\gamma$-ray variability can be counted as being detected.

\subsubsection{GJ 1078}
\label{sec:2.4}

For GJ~1078, there is no other contaminating sources and nearby 4FGL-DR3 sources (see Figure~\ref{fig:tsmap_2src}).
Thus, we directly use one point source (p4) with a power-law spectrum to model the excess to the southwest of GJ~1078 in the 1--500~GeV band and fit its position, yielding ${\rm R.A._{J2000}}=80.852^{\circ}$, ${\rm Dec_{J2000}}=22.333^{\circ}$ with a $1\sigma$ error radius of $0.066^{\circ}$.
The angular separation between p4 and GJ~1078 is $0.23^{\circ}$, which is greater than the 3$\sigma$ positional uncertainty ($0.15^{\circ}$) of p4.
So it is hard to associate p4 to GJ~1078, and thus no further analysis for p4 is performed.

\section{Discussion}
\label{sec:3}

\subsection{Physical association between p3 and V962\,Tau}

In our \textit{Fermi}-LAT data analysis described above for the eight red dwarfs, we find that only V962\,Tau may have the long-term GeV emission according to the possible spatial association between p3 and V962\,Tau. Here we discuss whether there may be a physical association between them in terms of radiation mechanism.

As one of the most studied objects, the Sun was observed by \textit{Fermi}-LAT in the past years \citep{Ajello2014,Ackermann2017}, showing hundreds of MeV to GeV $\gamma$-ray emission during some Solar flares.
The flares associated with $\gamma$-ray emission are generally accompanied by fast coronal mass ejections (CMEs).
In addition to the Sun, hard X-ray and/or soft $\gamma$-ray emission are also detected from stellar flares/superflares of nearby active stars, e.g., DG CVn \citep{Drake2014,2016hasa.book...23O}.
The magnetic reconnection and the CME-driven shock waves are invoked to explain the energetic particles which are responsible for the hard X-ray and/or $\gamma$-ray emission \citep{2022LRSP...19....2C}.
The former is generally related to the impulsive (prompt) phase of flares and is hard to accelerate particles beyond a few GeV \citep{Takahashi2016}.
After the peak of the impulsive phase, a longer duration phase (from several hours to weeks) of flares is followed, which can be attributed to the CME-driven shock waves.

\begin{figure*}
    \centering
	\includegraphics[width=8cm]{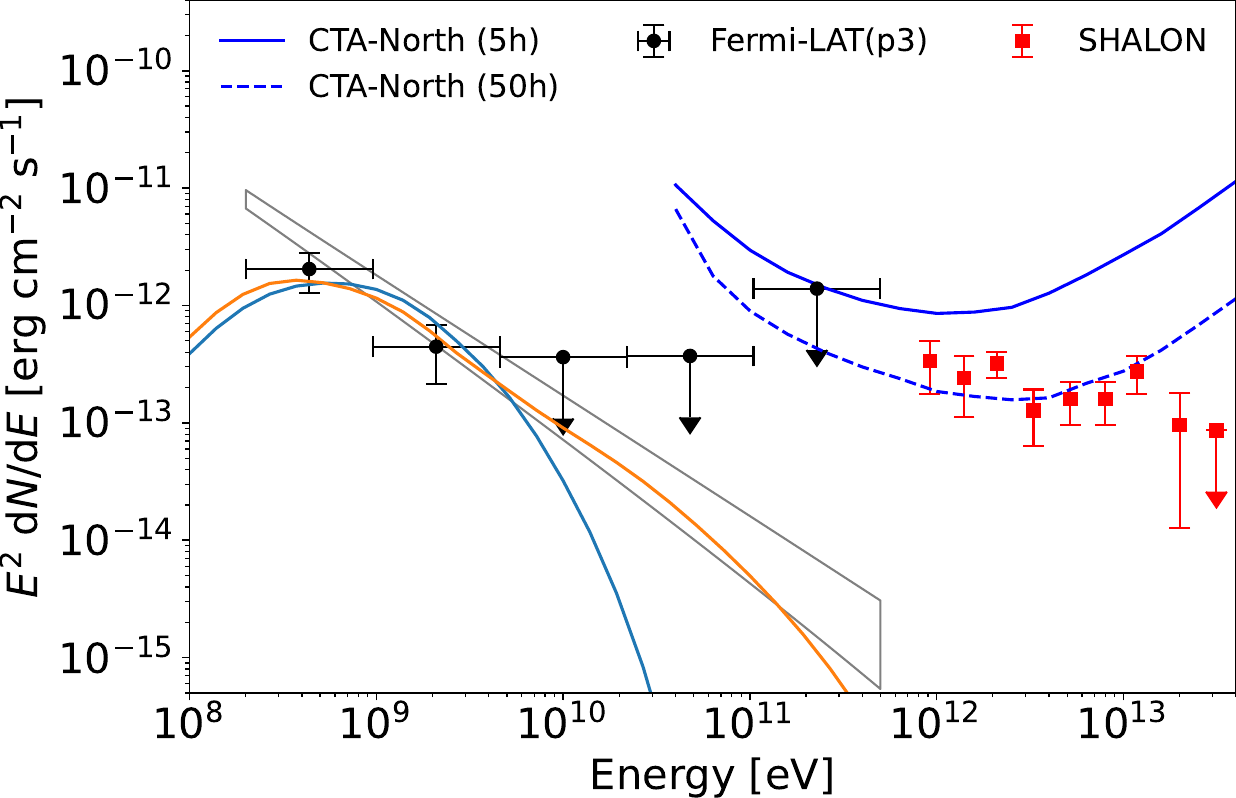}
  \caption{SEDs of p3 near V962~Tau. The black dots with the 1$\sigma$ error bars or the 95\% upper limits are \textit{Fermi}-LAT data obtained. The red squares with 1$\sigma$ error bars represent the TeV spectral data from SHALON toward these red dwarfs \citep{2021AN....342..342S}. The gray bowtie is the 68\% confidence range of the LAT spectra. The orange ($\alpha_{\rm p}=3.2$, $E_{\rm p,c}=1$~TeV, and $\eta=2.5$) and light blue ($\alpha_{\rm p}=2.0$, $E_{\rm p,c}=10$~GeV, and $\eta=1.5$) lines represent the fitting spectra of p3. The blue solid line and dashed line show the differential energy flux sensitivities of Cherenkov Telescope Array (CTA), for the exposure duration of 5~h and 50~h \citep{CTA_2021_5499840}.
  }
  \label{fig:sed}
\end{figure*}

Due to the very short duration of the impulsive phase, here we only adopt the CME-driven shock wave scenario, in which the shock accelerated protons with a fraction, $\eta$, of the CME kinetic energy collide with the CME matters to produce to pion-dacay $\gamma$-rays.
Taking the superflare of DG CVn as a prototype, \citet{10.1093/mnras/stx2806} calculated the expected $\gamma$-ray emission from the nearby flaring stars.
Following their treatment, for the CME kinetic energy, we take an optimistic value $E_{\rm CME,kin}=10^{37}$~erg as a reference, which corresponds to the CME mass of $2.2\times10^{20}$~g and the CME velocity of $3000\ {\rm km\ s^{-1}}$.
With these parameters, the average density of the CME particles as the target of the relativistic protons is $n_{11}=2$ in unit of $10^{11}\ {\rm cm^{-3}}$ in a few hours of evolution after the flaring.
The accelerated protons are assumed to have an exponentially cut-off power-law spectrum with an index of $\alpha_{\rm p}$ and a cutoff energy $E_{\rm p,c}$.
We adopt the python package Naima\footnote{\url{https://naima.readthedocs.io/en/latest/mcmc.html}} (version 0.10.0) with the proton-proton cross section from \citet{2014PhRvD..90l3014K} to calculate the pion-decay $\gamma$-rays.
In the calculation, we found that the parameters are degenerated and cannot be constrained by the current data points.
Thus, we consider two cases: 1) the index is fixed as 2.0 which is the typical value predicted by the shock acceleration theory; and 2) the cutoff energy is fixed as 1 TeV which is the possible maximum energy in the flares \citep{10.1093/mnras/stx2806}.
At a distance of 710~pc \citep{2020yCat.1350....0G}, we obtain $E_{\rm p,c}=10$~GeV, and $\eta=1.5$ in the former case and $\alpha_{\rm p}=3.2$ and $\eta=2.5$ in the latter to explain the \textit{Fermi}-LAT data of source p3.
The corresponding SED is presented in Figure~\ref{fig:sed}.

On one hand, even with the optimistic CME kinetic energy, the energy conversion fraction $\eta>1$ indicates that one flare can not alone maintain the GeV $\gamma$-ray luminosity of p3.
Giving the short lifetime of the accelerated protons $2.6n_{11}^{-1}$~h \citep{10.1093/mnras/stx2806} and an optimistic assumption that the CME can transfer 10\% of its total energy to accelerate particles and all of them interact with the surrounding medium for $\gamma$-ray emission in a duration comparable to the proton-proton loss time, fifteen superflares at least are needed to produce the observing flux.
However, V962 Tau has the largest distance in our sample. If it has a few tens parsec like other dwarfs, the energy budget can decrease three orders of magnitude, which could be supplied by the CME kinetic energy.
On the other hand, considering low variability of LC of p3, the superflares should occur very frequently, which seems unreasonable according to the current statistics of flares \citep{2017ApJ...834...92C, 2019ApJS..241...29Y}.
Thus, it is difficult for source p3 to be physically related to the flares from V962\,Tau.

Besides the above flare scenario, the red dwarfs could be potential passive $\gamma$-ray source: the background CRs could penetrate into the atmosphere of them and emanate $\gamma$-ray emission.
Following the work for the case of the Sun \citep{2017PhRvD..96b3015Z}, the passive $\gamma$-ray flux from V962\,Tau is estimated as $\sim10^{-29}$ ${\rm erg\ cm^{-2}\ s^{-1}}$, far below the flux of p3 ($\sim10^{-13}$ ${\rm erg\ cm^{-2}\ s^{-1}}$).
Thus the passive $\gamma$-ray emitter scenario for p3 can be excluded.
In addition, we search for other possible counterparts within $3\sigma$ error radius of p3 in SIMBAD \citep{2000A&AS..143....9W} and find that there is no source belonging to the types of known $\gamma$-ray-bright sources.
Due to the low significance of $\sim4\sigma$ and the low Galactic latitude $|b|\sim3^{\circ}$, p3 appearing near V962\,Tau may likely be a background emission in nature.

\subsection{Constraints for upper limits}
\label{sec:3.2}

Based on the above analysis and discussion, in fact, there is no long-term GeV $\gamma$-ray emission for eight red dwarfs including V962 Tau.
So we calculate the upper limits to constrain some physical properties.
Assuming a point source and subtracting the nearby excess for each source, the upper limits are calculated in five energy bins based on a Power-Law spectrum with a spectral index of 2.0 in each bin, which are displayed in Figure~\ref{fig:sed_8}.
With $E_{\rm CME,kin}=10^{37}$~erg, $\alpha_{\rm p}=2.0$, and $E_{\rm p,c}=10$~GeV, the energy conversion fraction can be constrained to $\eta \lesssim 10^{-4}$.
In the paradigm of supernova remnants as the main sources of Galactic cosmic rays, the particles accelerated by the shocked should carry 10\% of the explosion energy \citep[e.g.,][and references therein]{blasi2013}.
If the CME shocks can also accelerate particles with the same energy conversion fraction, then the kinetic energy of CMEs can be reduced to the order of $10^{34}$ erg.
For flares with such energy, the frequency only can reach several times per one year.
Thus, the long-term GeV $\gamma$-ray emission from red dwarfs may be hardly to detected by the current telescope.

With an the occurrence frequency of stellar flares ($\nu\sim 36\ {\rm yr^{-1}}$) and the average flare energy ($\sim10^{35}$ erg), it was suggested that flares of active dwarf stars can also provide enough energy to maintain the luminocity of the Galactic CRs \citep{Kopysov2005}.
The energy density of the Galactic CRs is about $w \sim1$ eV cm$^{-3}$ \citep[e.g.,][]{gaisser1990}.
By adopting a radius of 10 kpc and a thickness of 300 pc, the volume of the Galactic disk is $V_{\rm gal}\sim3\times10^{66}\ {\rm cm}^{-3}$.
Considering the lifetime of CRs in the disk $\tau\sim10^{7}$ yr, the luminosity of the Galactic CRs is $L_{\rm}=wV_{\rm gal}/\tau\sim 1.4\times 10^{40}\ {\rm erg\ s^{-1}}$.
In Galaxy, there are $n_{\rm s}\sim 2\times10^{11}$ stars, and most of them are the G-M spectral classes.
Assuming the accelerated particles taking up 10\% of the flare energy, the power of CRs contributed by flares of the active dwarf stars is $0.1n_{\rm s}\nu E_{\rm CME,kin}$.
To explain the luminosity of CRs, $\nu E_{\rm CME,kin} \sim 1.4\times 10^{30}\ {\rm erg\ s^{-1}} (n_{\rm s}/10^{11})^{-1}$ is required.
According to our constraints $E_{\rm CME,kin}<10^{34}\ {\rm erg}$ and the occurrence frequency of such flares $\nu < 10\ {\rm yr^{-1}}$ \citep[e.g.,][]{2019ApJS..241...29Y}, flares of the active dwarf stars only has a limited contribution to the Galactic CRs.

In addition, if the particle index $\alpha_{\rm p}>2.0$ and the maximum energy $E_{\rm p,c}>\sim10$~TeV, the SHALON data are obviously higher than the theoretical expectation constrained by the Fermi-LAT upper limits.
This means that the TeV emission seen by SHALON maybe not associate with red dwarfs, which can be verified by the next generation telescope, e.g., CTA.

\subsection{Flare-like events}
Although no long-term GeV $\gamma$-ray emission is found for the eight red dwarfs, there may be flares in the GeV band like the case of the Sun reflected in the LC.
Based on the best-fit source model derived from the binned likelihood analysis in Section~\ref{sec:2}, we construct the one-week binned LC for each red dwarf over the entire data.
For V962~Tau, V1589~Cyg, and GJ~1078, surrounding excesses are included in the source model and are treated as background sources.
We first search for the flare-like candidate events via seeking the time intervals in which the $\gamma$-ray emission is detected with large significance (e.g., TS $\geq$ 25).
But no such intervals are found for all the eight red dwarfs.
Due to the blind search, the period of a flare may have been divided into two time bins, reducing the significance of the $\gamma$-ray signal.
We thus deletes the first 3.5 d data and re-built the one-week binned LCs over the remaining data, but, again, no periods with significant $\gamma$-ray signal are found.
Thus, we conclude that no flare is detected by \textit{Fermi}-LAT for the eight red dwarfs based on the current observational data.

\section{Summary}
\label{sec:4}
In order to search for potential GeV $\gamma$-ray emission from red dwarfs, we analyse the 0.2--500\,GeV $\gamma$-ray emission of the regions covering V388~Cas, V547~Cas, V780~Tau, V962\,Tau, V1589~Cyg, GJ\,1078, GJ~3684 and GL~851.1 with 13.6 yr \textit{Fermi}-LAT data and find significant excesses projected near V962\,Tau and GJ\,1078. The spatial analysis shows that V962\,Tau is in the range of 1$\sigma$ error radius of the nearby $\gamma$-ray emission after taking the systematic error into consideration, while GJ\,1078 are out of the 3$\sigma$ error radius of nearby emission, meaning that it is hard to associate them with the nearby excess. The spectral analysis shows that the $\gamma$-ray emission near V962\,Tau can be fitted by exponentially cut-off power-law proton distribution models.
However, on the optimistic assumption of CME kinetic energy ($10^{37}$~erg) we obtain the energy conversion efficiency $\eta>1$, which is impossible for a single flare. Considering the low variability of LC of p3, the flares should be very frequent, which is unreasonable according to the statistics of flare. As a result, the GeV $\gamma$-ray emission presented should be irrelevant with V962\,Tau. We also construct the one-week binned LC for all of the eight red dwarfs but no periods with significant emission are found.
Therefore, more observation about high energy processes is still necessary to explore nearby flaring stars and is expected to increase our understanding of the physics of stellar energetic eruption events.


\begin{acknowledgments}
X.Z.\ and Y.C.\ thank the support of National Key R\&D Program of China under no.\ 2018YFA0404204 
and NSFC grants under nos.\ 12393852, U1931204, 11803011, 12173018, and 12121003. This research has made prominent use of the SIMBAD data base.
\end{acknowledgments}

\appendix
\section{Some extra material}
\label{appendix}
\begin{figure*}
    \centering
        \includegraphics[width=8cm]{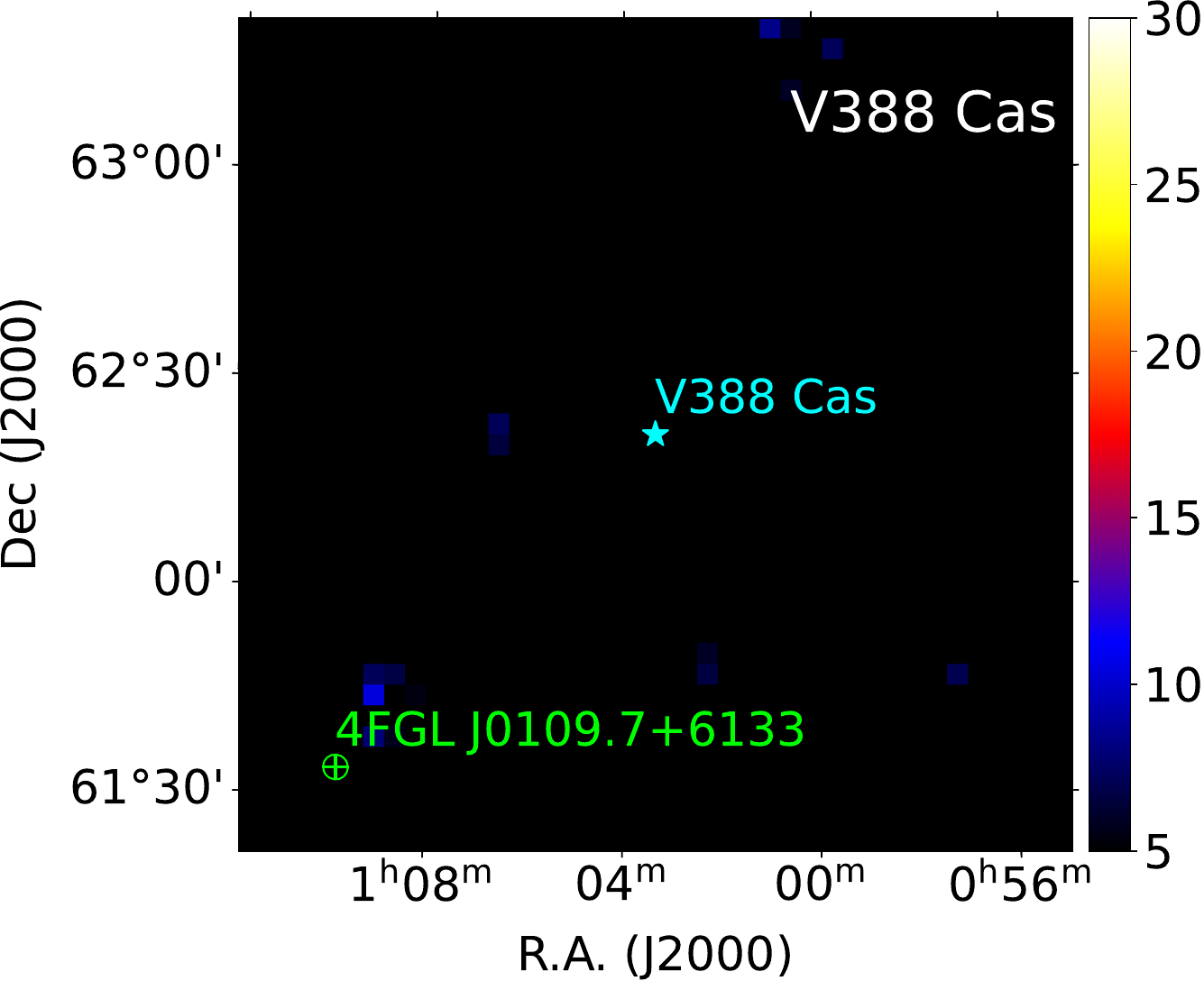}
        \includegraphics[width=8cm]{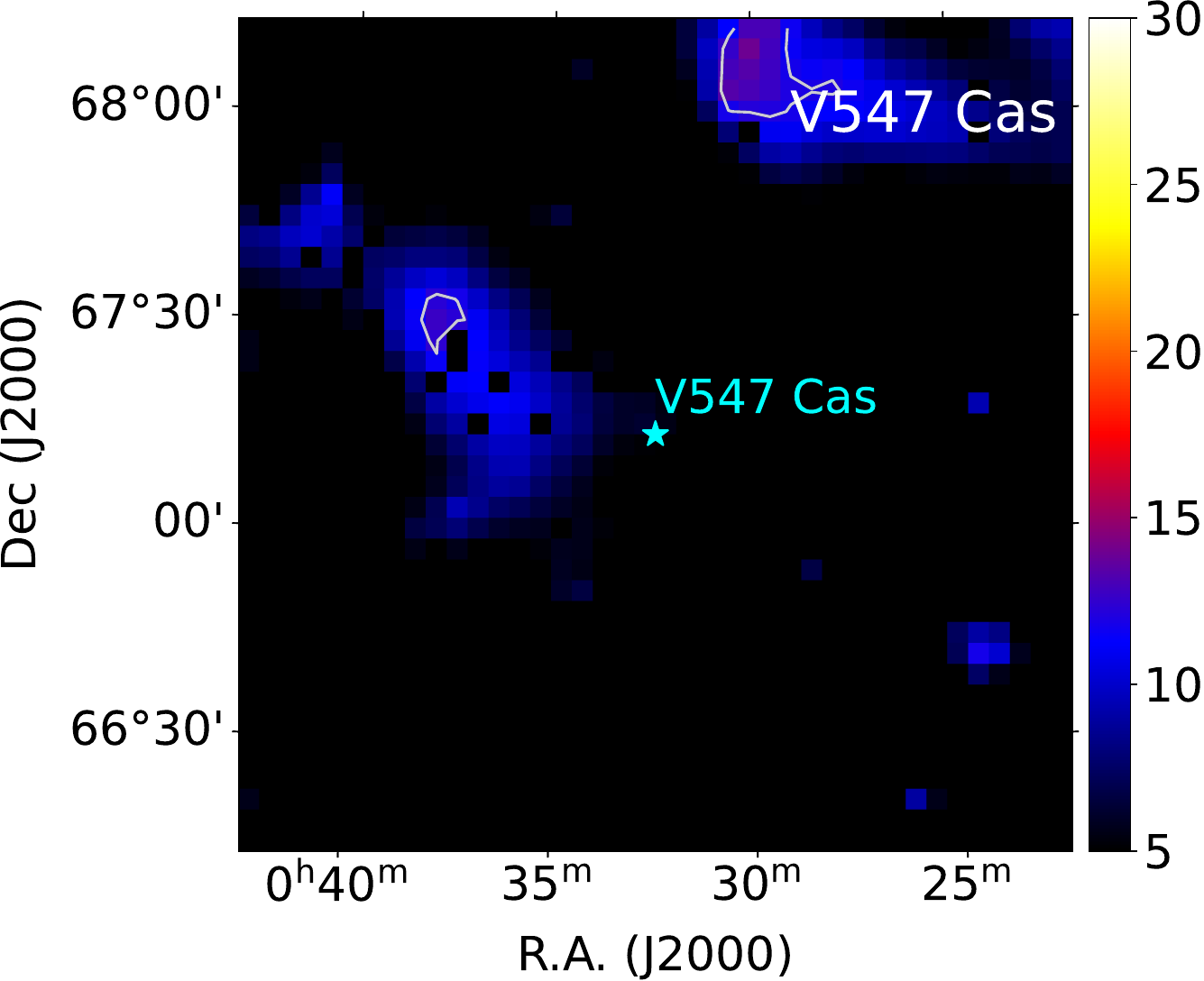}
        \includegraphics[width=8cm]{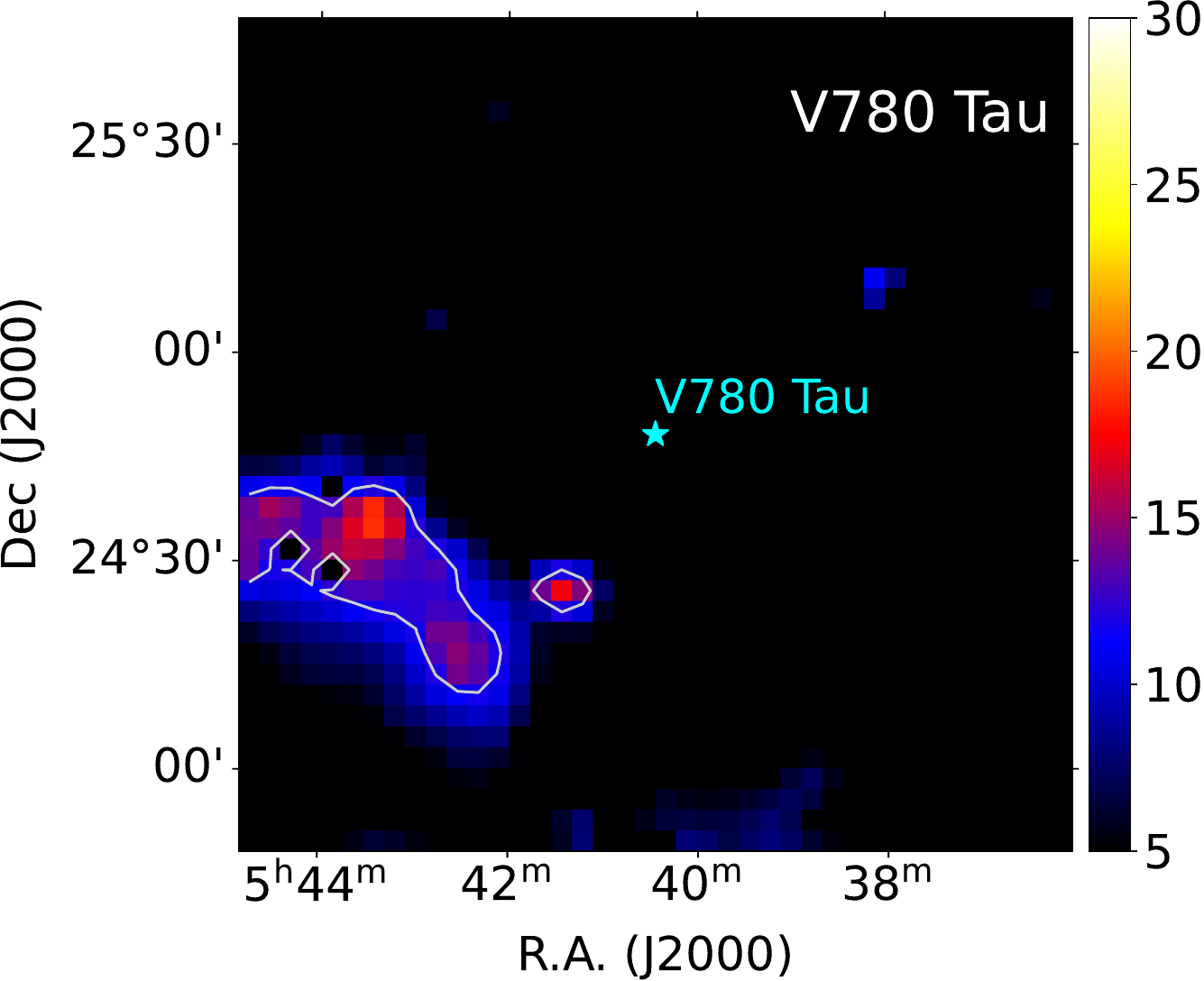}
        \includegraphics[width=8cm]{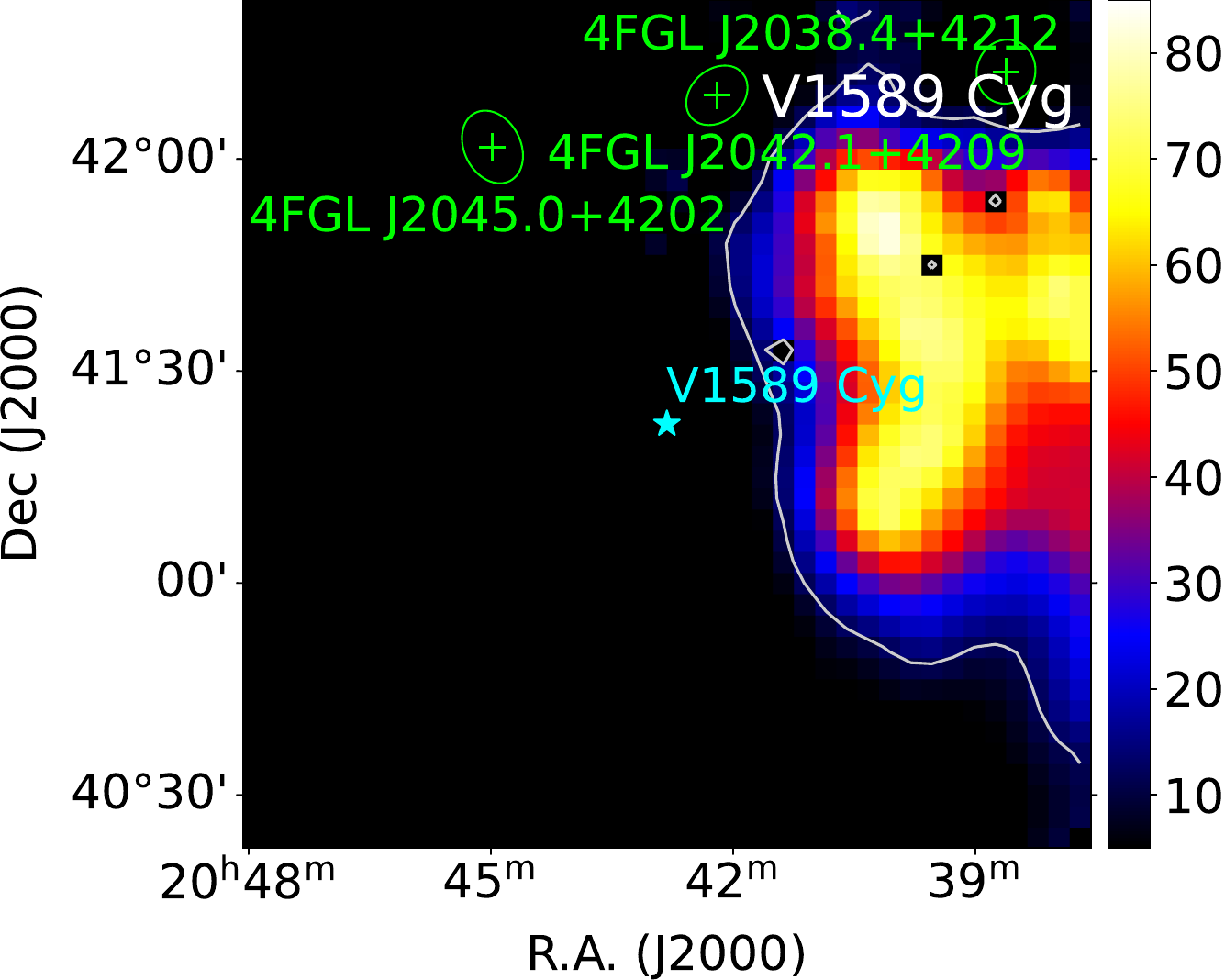}
        \includegraphics[width=8cm]{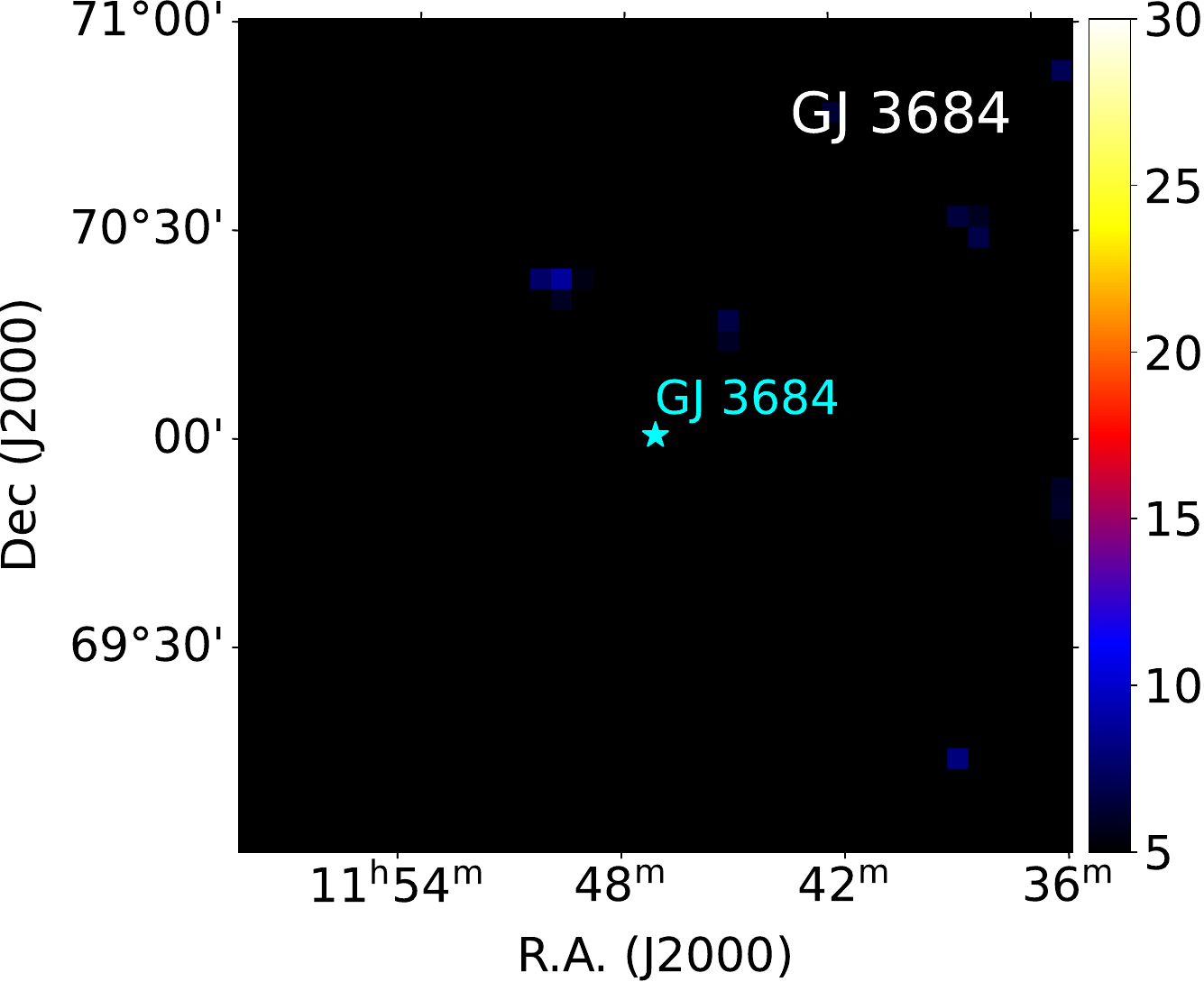}
        \includegraphics[width=8cm]{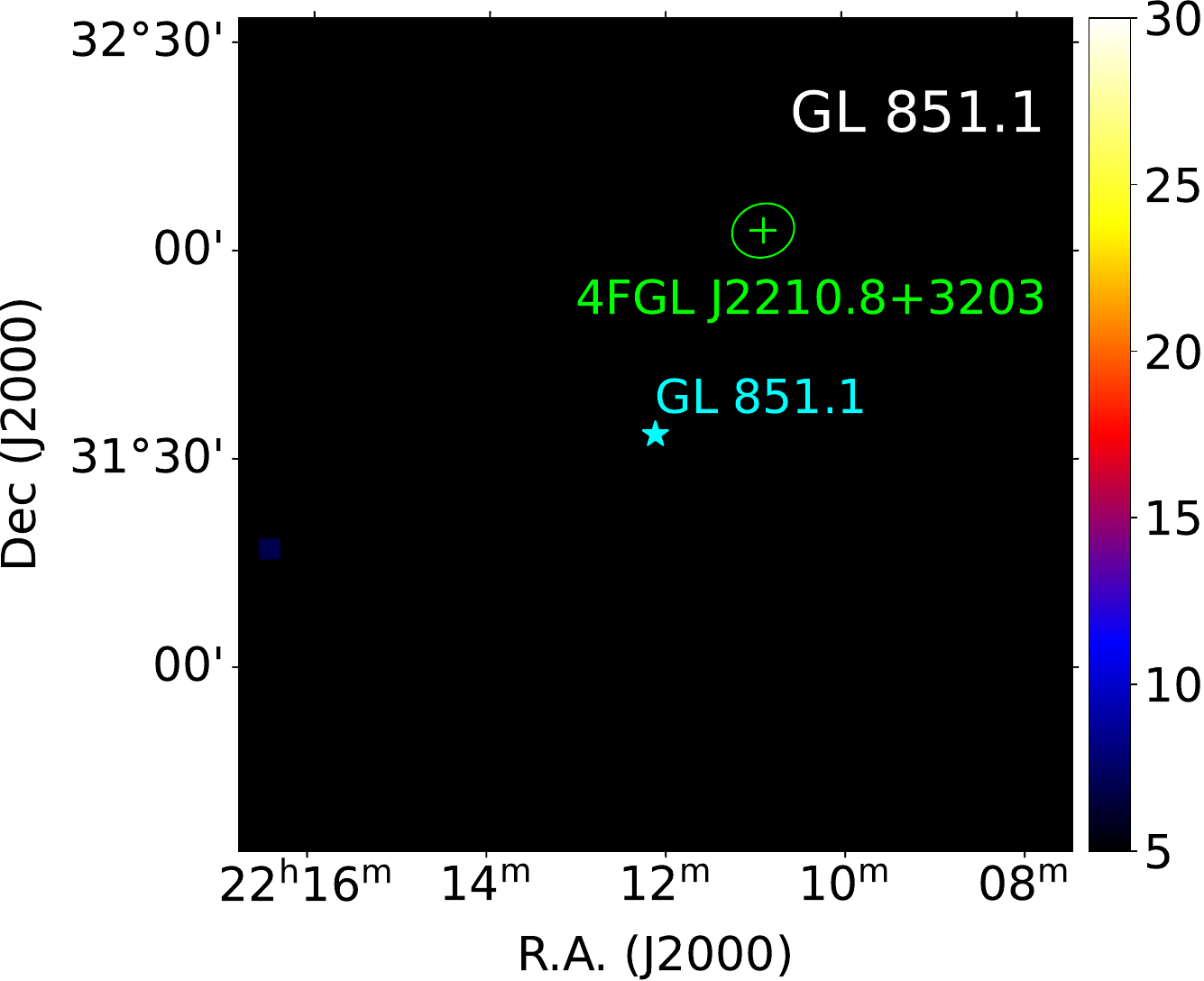}
    \caption{TS maps of 2$^{\circ}$ × 2$^{\circ}$ regions centered at V388~Cas,V547~Cas,V780~Tau, V1589~Cyg, GJ~3684 and GL~851.1 in the energy range of 1--500~GeV.}
    \label{fig:tsmap_6src}
\end{figure*}

\begin{figure*}
    \centering
    \includegraphics[width=8cm]{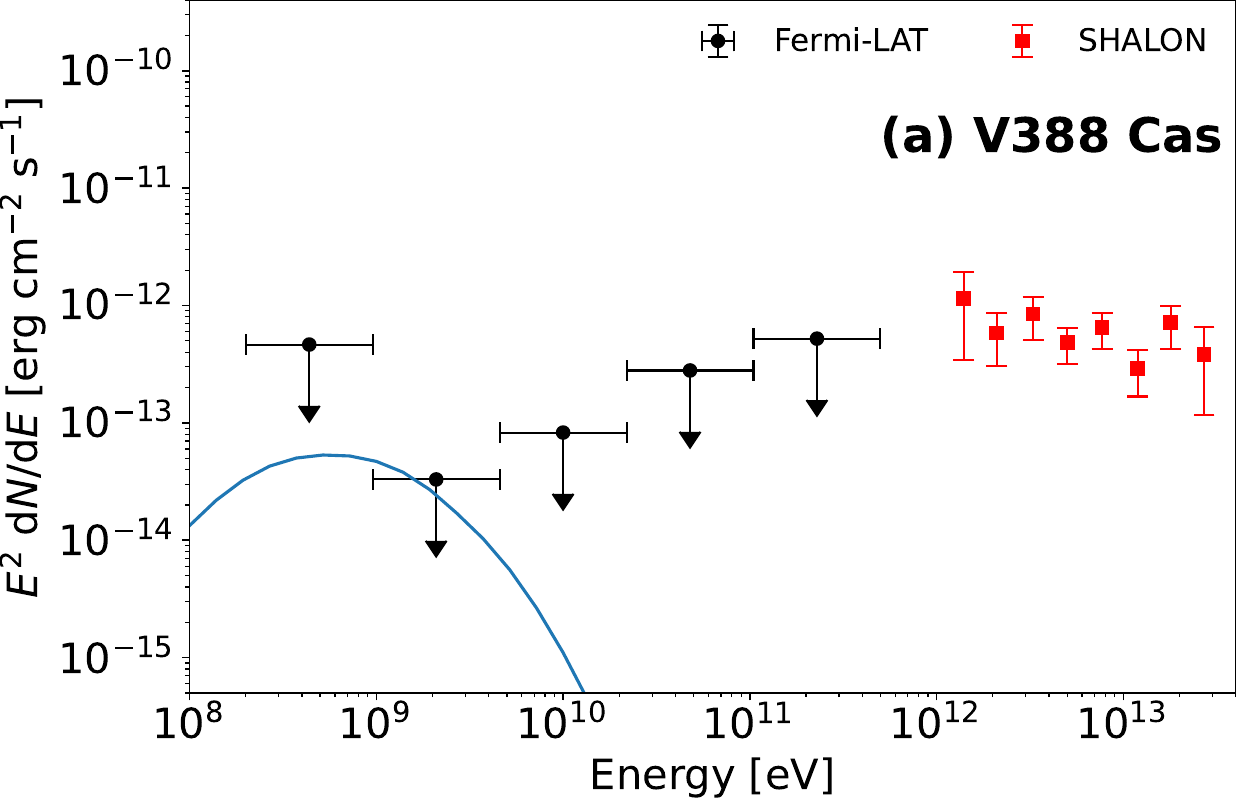}
    \includegraphics[width=8cm]{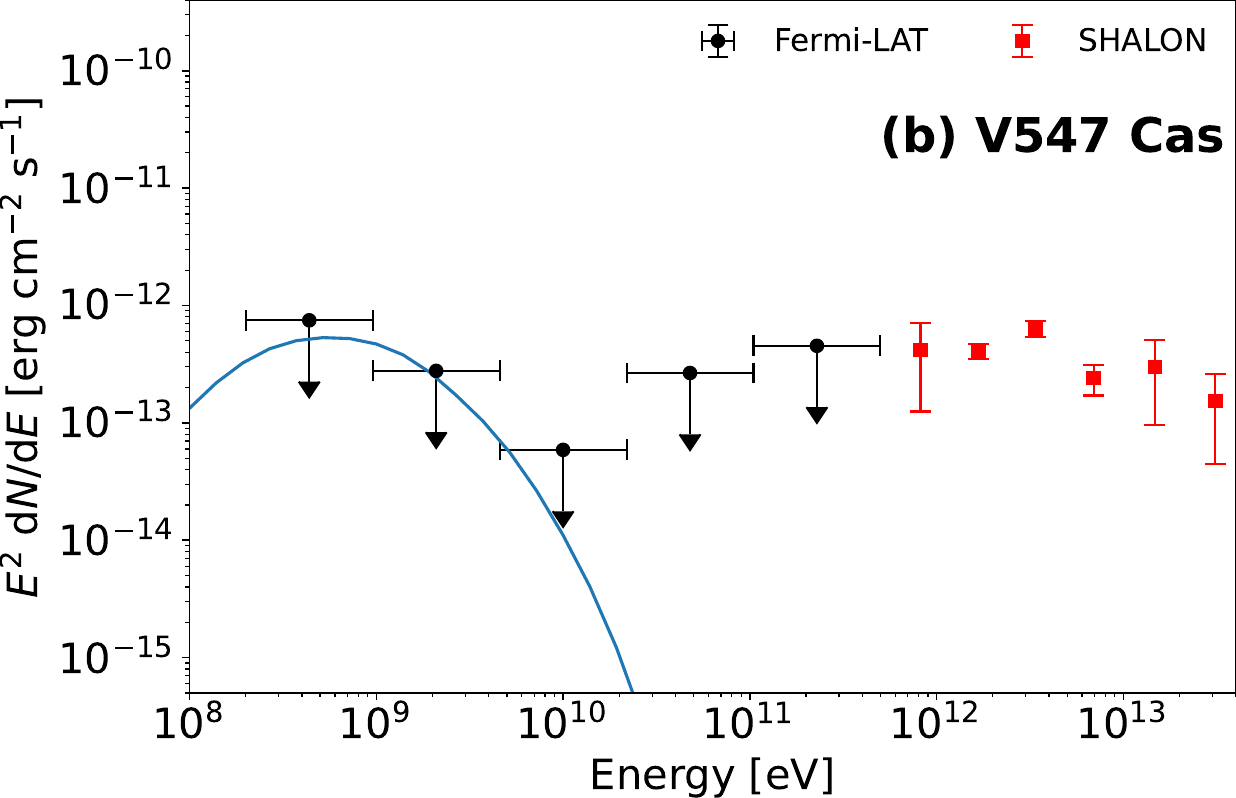}
    \includegraphics[width=8cm]{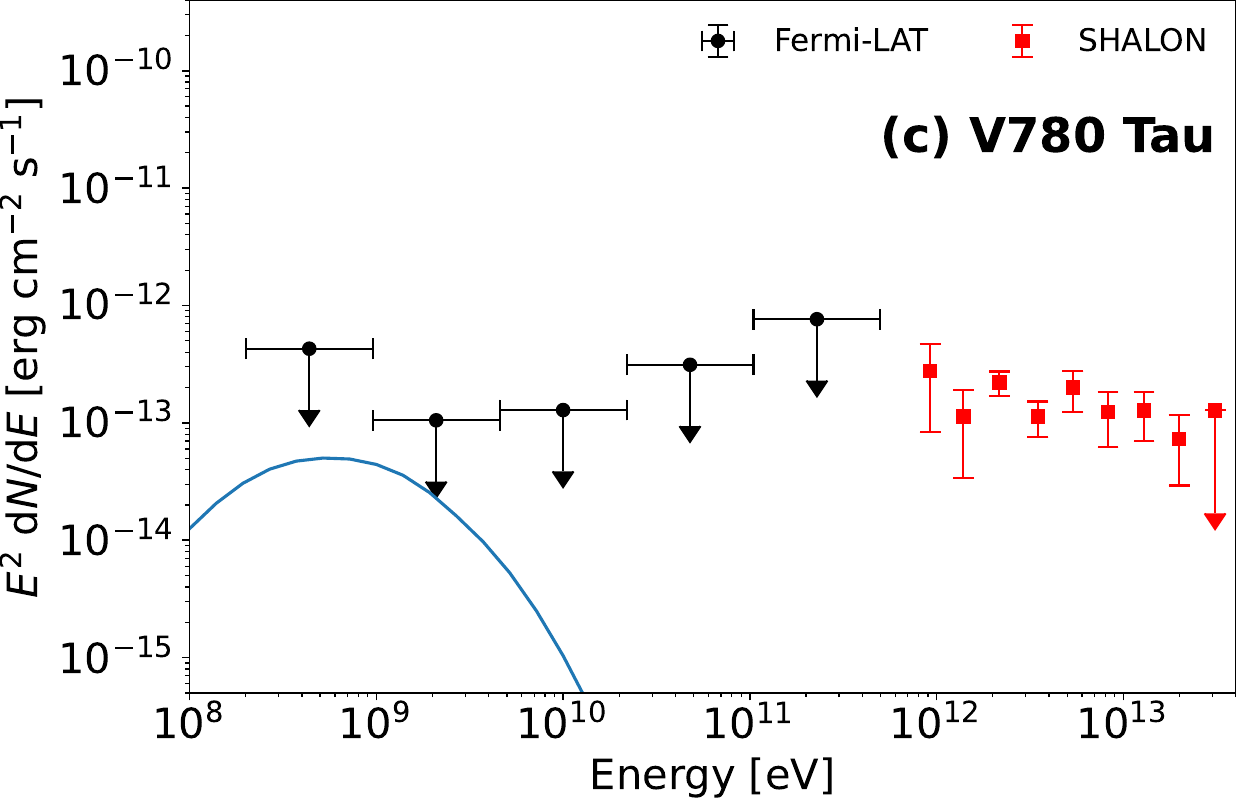}
    \includegraphics[width=8cm]{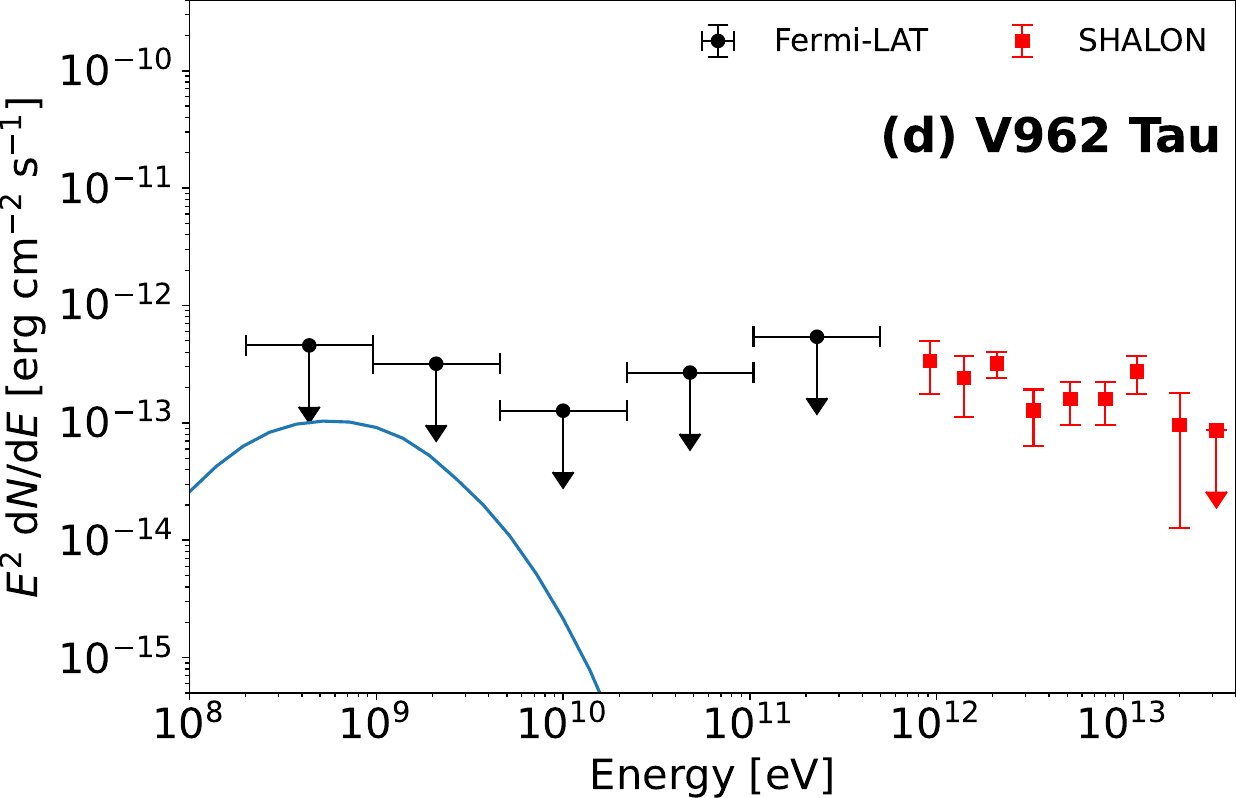}
    \includegraphics[width=8cm]{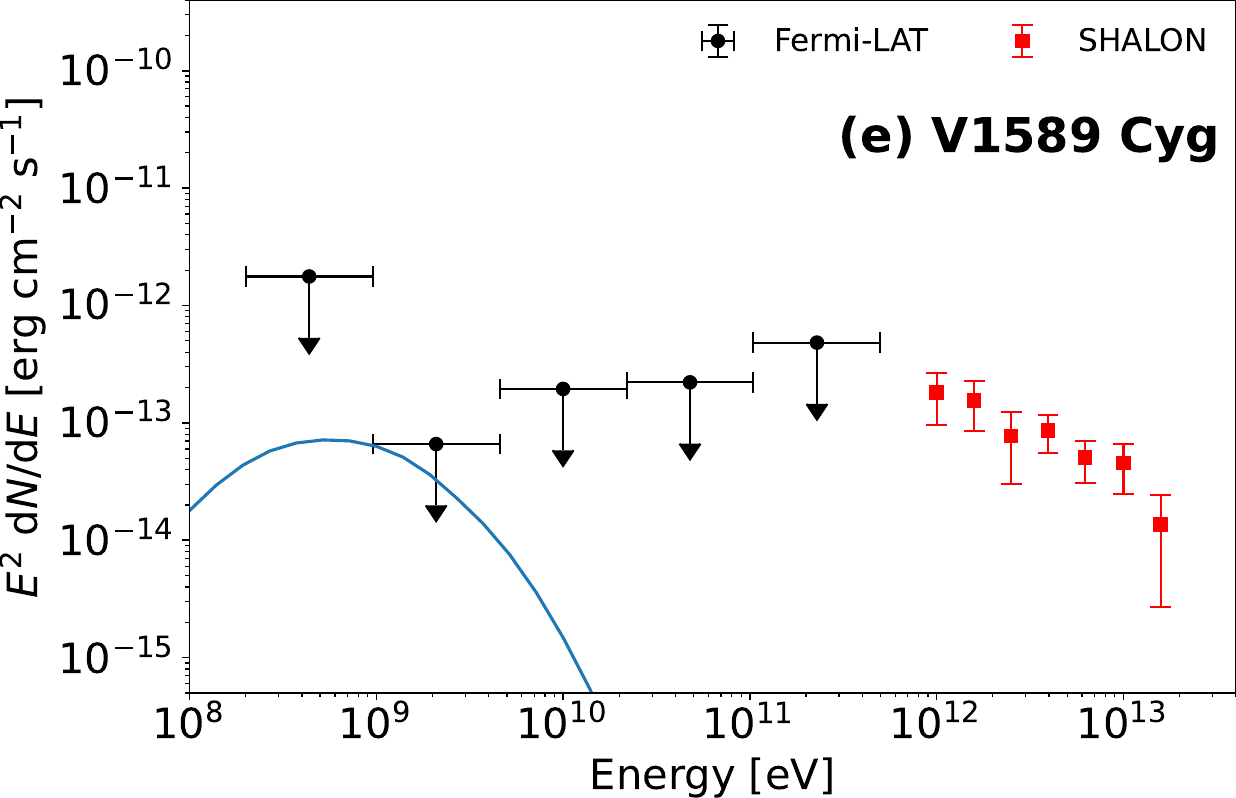}
    \includegraphics[width=8cm]{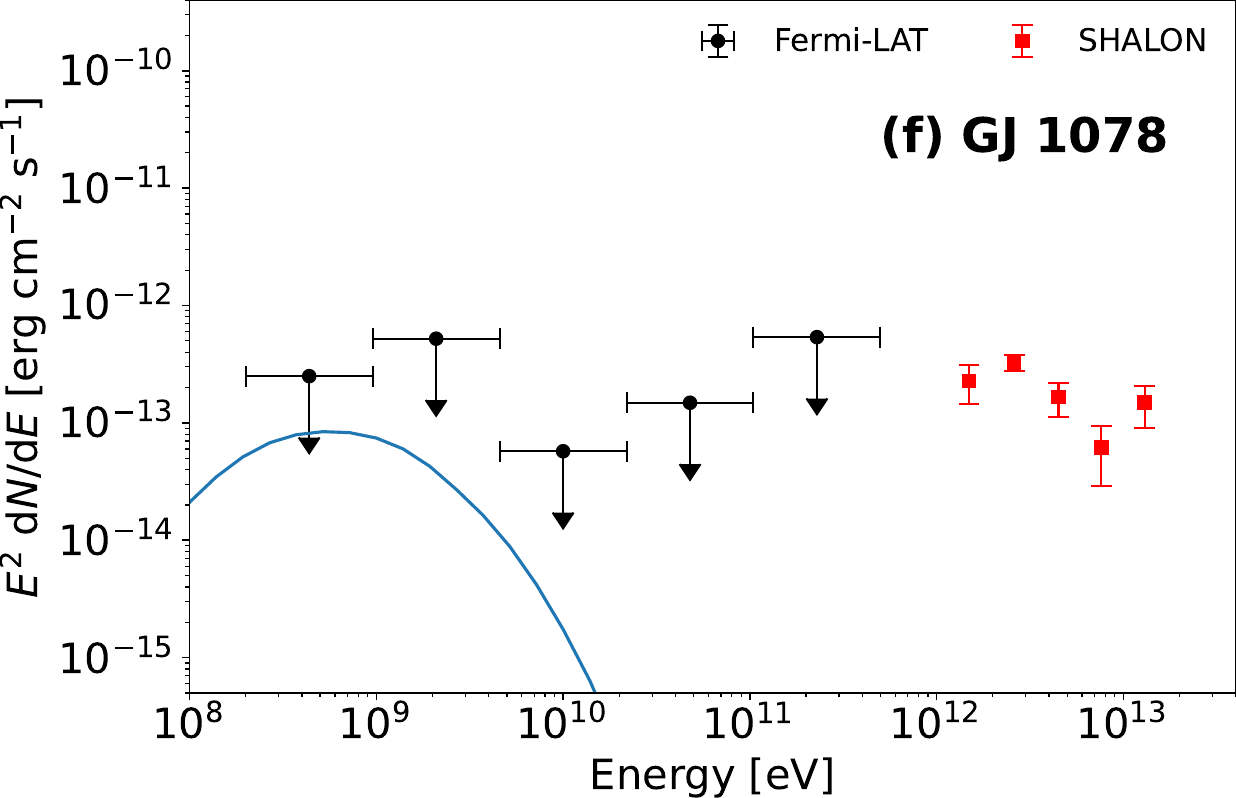}
    \includegraphics[width=8cm]{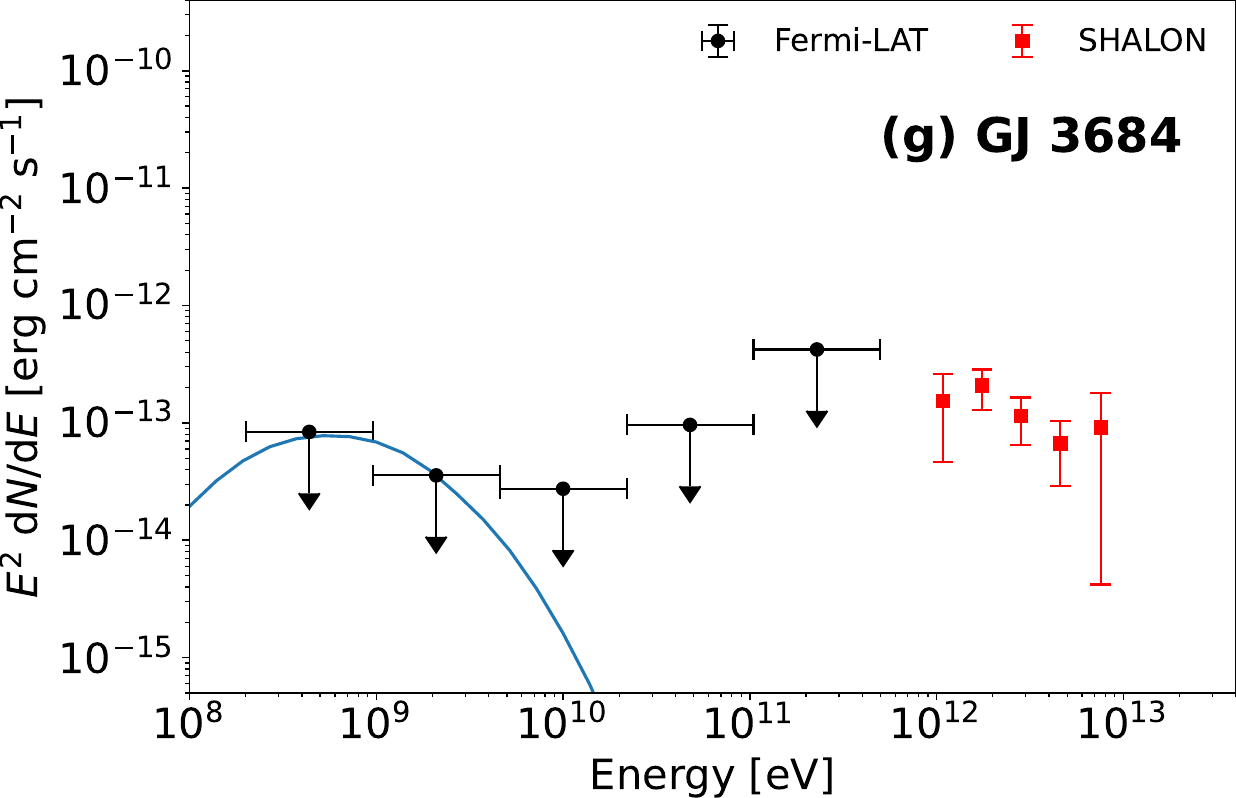}
    \includegraphics[width=8cm]{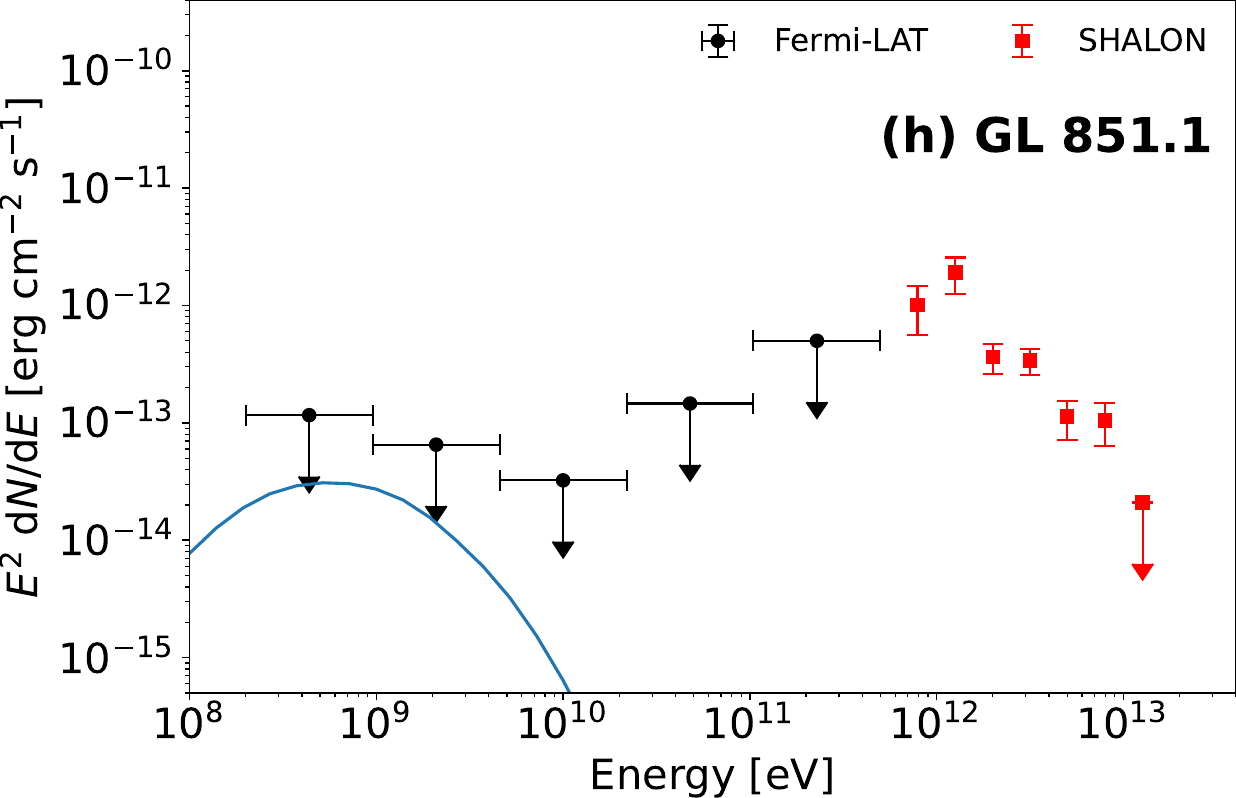}
    
  \caption{Same as Figure~\ref{fig:sed} but for (a) V388~Cas, (b) V547~Cas, (c) V780~Tau, (d) V962~Tau, (e) V1589~Cyg, (f) GJ~1078, (g) GJ~3684, and (h) GL~851.1. The light blue lines show the pion-decay $\gamma$-rays with parameters of $\alpha_{\rm p}=2.0$, $E_{\rm p,c}=10$~GeV, and $\eta=10^{-4}$ ($10^{-1}$ for V962~Tau).
  }
  \label{fig:sed_8}
\end{figure*}

\restartappendixnumbering
We also plot the TS maps of the other sources as reference, which are shown in Figure~\ref{fig:tsmap_6src}. Moreover, SEDs of all the eight red dwarfs are shown in Figure~\ref{fig:sed_8}, with elaboration in Section~\ref{sec:3.2}.

\bibliography{references}{}
\bibliographystyle{aasjournal}

\end{document}